\documentclass{nature_mod}
\usepackage{graphicx}
\usepackage{graphics}
\usepackage{ulem}

\usepackage{amsmath,amssymb}

\newcommand{\apj}{Astrophys.~J.}      
\newcommand{\aj}{Astron.~J.}      
\newcommand{\aap}{Astron.~Astrophys.}  
\newcommand{\apjl}{Astrophys.~J.~Letters}   
\newcommand{\mnras}{Mon.~Not.~R.~Astron.~Soc.} 

\newcommand{\pasj}{Pub.~Astron.~Soc.~Japan}   
\newcommand{\araa}{Ann.~Rev.~Astron.~Astrophys} 
\newcommand{\bain}{Bull.~Astron.~Inst.~Netherlands} 

\title{Interacting supernovae from photoionization-confined shells around red supergiant stars}

\author{Jonathan Mackey$^{1}$, Shazrene Mohamed$^2$, Vasilii V.\ Gvaramadze$^{3,4,5}$, Rubina Kotak$^6$, Norbert Langer$^{1}$,  Dominique M.-A.\ Meyer$^{1}$, Takashi  J.\ Moriya$^{1}$ \& Hilding R.\ Neilson$^{7}$}

\begin{document}

\maketitle

\begin{affiliations}
 \item Argelander-Institut f\"ur Astronomie, Auf dem H\"ugel 71, 53121 Bonn, Germany
 \item South African Astronomical Observatory, P.O.\ box 9, 7935 Observatory, South Africa
 \item Sternberg Astronomical Institute, Lomonosov Moscow State University, Universitetskij Pr.~13, Moscow 119992, Russia
 \item Isaac Newton Institute of Chile, Moscow Branch, Universitetskij Pr.\ 13, Moscow 119992, Russia
 \item Space Research Institute, Russian Academy of Sciences, Profsoyuznaya 84/32, 117997 Moscow, Russia
 \item Astrophysics Research Centre, School of Mathematics and Physics, Queen's University Belfast, Belfast BT7 1NN, UK
 \item Department of Physics and Astronomy, East Tennessee State University, Box 70652, Johnson City, TN, 37614, USA
\end{affiliations}

\begin{abstract}
Betelgeuse, a nearby red supergiant, is a runaway star with a powerful stellar wind that drives a bow shock into its surroundings\cite{NorBurCaoEA97, UetIzuYamEA08, MohMacLan12, DecCoxRoyEA12}.
This picture has been challenged by the discovery of a dense and almost static  shell\cite{LeBMatGerEA12} that is three times closer to the star than the bow shock and has been decelerated by some external force.
The two physically distinct structures cannot both be formed by the hydrodynamic interaction of the wind with the interstellar medium.
Here we report that a model in which Betelgeuse's wind is photoionized by radiation from external sources can explain the static shell without requiring a new understanding of the bow shock.
Pressure from the photoionized wind generates a standing shock in the neutral part of the wind\cite{Kah54} and forms an almost static, photoionization-confined shell.
Other red supergiants should have significantly more massive shells than Betelgeuse, because
the photoionization-confined shell traps up to 35 per cent of all mass lost during the red supergiant phase, confining this gas close to the star until it explodes.
After the supernova explosion, massive shells dramatically affect the supernova lightcurve, providing a natural explanation for the many supernovae that have signatures of circumstellar interaction.
\end{abstract}

Red supergiants are massive stars near the end of their lives, and are direct progenitors of core-collapse supernovae\cite{Sma09, Lan12}.
They evolve from O- and B-type stars (hot, luminous sources of ionizing photons), and so these stars are often found together, within or near star clusters\cite{DouClaNegEA10}.
As a result, the cool stellar winds of red supergiants are often photoionized by external radiation fields\cite{MorJur83, YusMor91, WriWesDreEA14, GvaMenKniEA14}. 
To calculate the radiation hydrodynamics of a photoionized red supergiant wind, we simplify the problem by assuming spherical symmetry.
We use an approximate two-temperature equation of state for the gas, for which both the neutral and photoionized gases are isothermal with temperatures $T=T_\mathrm{n}$ and $T_\mathrm{i} \gg T_\mathrm{n}$, respectively.
The ionized and neutral isothermal sound speeds similarly satisfy $a_\mathrm{i} \gg a_\mathrm{n}$.
The photoionized part of the red supergiant wind is accelerated as a result of ionization heating\cite{MeyGvaLanEA14},
 whereas the neutral part is decelerated\cite{Kah54} if the wind speed through the ionization front, $v_\mathrm{n}$, satisfies $v_\mathrm{n}\leq2a_\mathrm{i}$.

\begin{figure}
\caption{
  \textbf{
  Circumstellar structures produced when a runaway red supergiant is exposed to an external ionizing radiation field.}
  {\bf [Left]}  A neutral stellar wind expands freely from the star and is shocked and decelerated by a photoionization-confined shell.
  A photoionized wind accelerates away from the shell's outer surface until it reaches the interface between the wind and the interstellar medium, which is a bow shock for Betelgeuse.
  {\bf [Right]} Detailed structure of a photoionization-confined shell from a spherically symmetric radiation hydrodynamics simulation of Betelgeuse's wind.
  Hydrogen number density, $n_\mathrm{H}$, velocity, $v$, and temperature, $T$, are plotted as functions of radius.   \\ .
  \label{fig:cartoon}
  }
\begin{center}
\includegraphics[width=0.49\hsize]{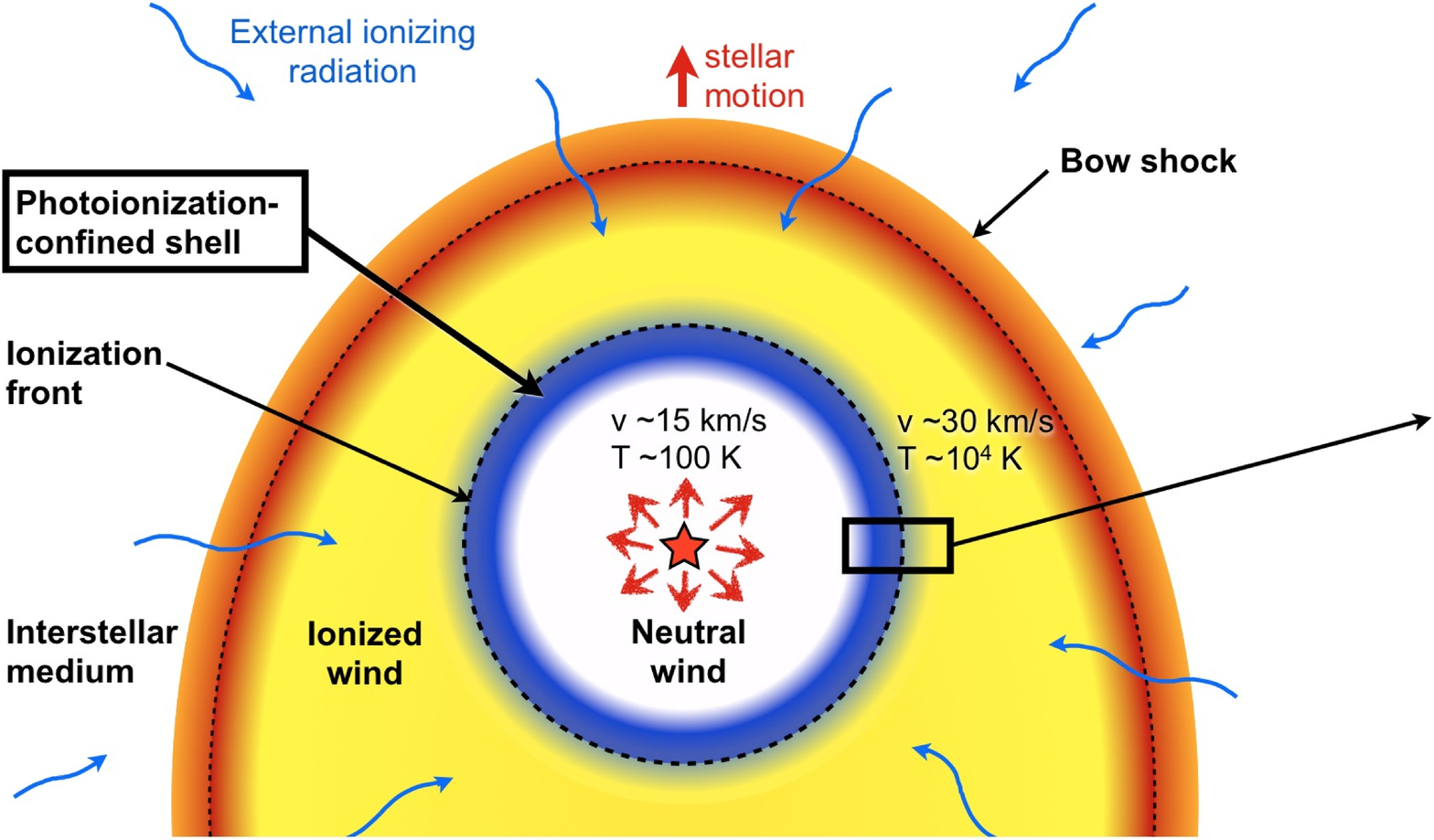}
\includegraphics[width=0.49\hsize]{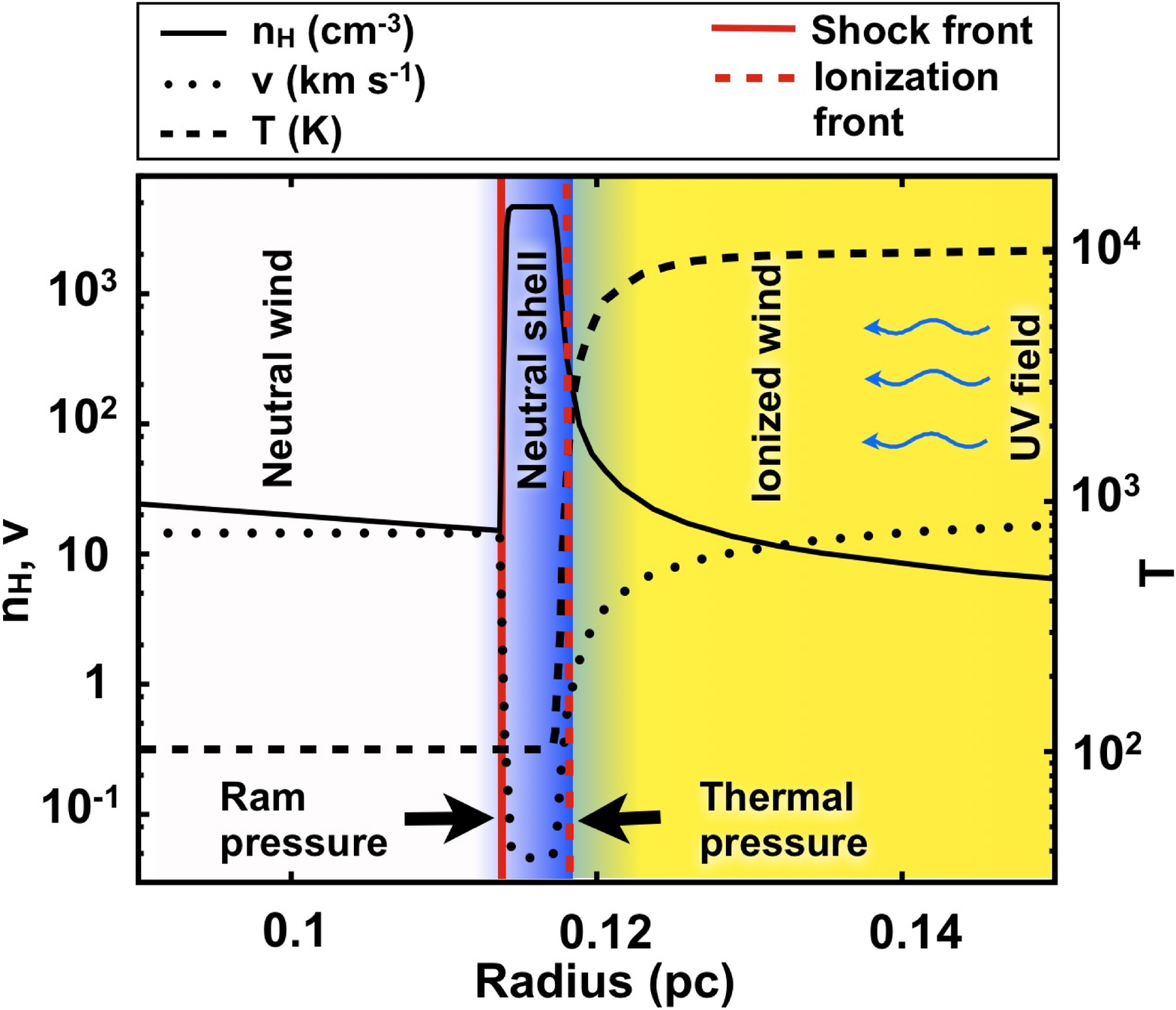}
\end{center}
\end{figure}

The resulting flow is depicted in Fig.~\ref{fig:cartoon}.
The outermost layer is the interface where the wind meets the interstellar medium.
For static stars this is a spherical, detached shell, and for stars moving supersonically it is a bow shock.
A photoionization-confined shell -- a dense, shocked layer separating the neutral inner wind from the ionized outer wind -- forms closer to the star.
We identify this with the recently-discovered shell in Betelgeuse's circumstellar medium\cite{LeBMatGerEA12}.

The properties of the photoionization-confined shell are calculated analytically and verified with simulations in Methods.
Its outer boundary, $R_\mathrm{IF}$, is calculated following previous work\cite{MorJur83} (Extended Data Fig.~1), and the standing shock radius, $R_{\mathrm{shell}}$, is obtained by requiring pressure balance across the shell.
 The shell reaches its final position (determined by the wind density and the incoming photon flux) on the expansion timescale of the wind, and then accumulates mass until it reaches a steady state, where the gas added to the shell at $R_\mathrm{shell}$ is balanced by that photoevaporated from $R_\mathrm{IF}$.
The steady-state mass of the shell, $M_\mathrm{shell}$, follows from its density and volume (Extended Data Fig.~2).
For realistic wind properties and radiation fluxes, the most likely radii and masses are $R_\mathrm{IF}\approx(0.003-0.3)$ pc and  $M_\mathrm{shell}\approx (0.03-10)\ M_{\odot}$ ($M_\odot$, solar mass).
Photoionization-confined shells are present \textit{in addition} to bow shocks and detached shells, and should be common because red supergiant winds are often photoionized\cite{MorJur83, YusMor91, WriWesDreEA14, GvaMenKniEA14}.

\begin{figure}
\begin{center}
\includegraphics[width=0.48\hsize]{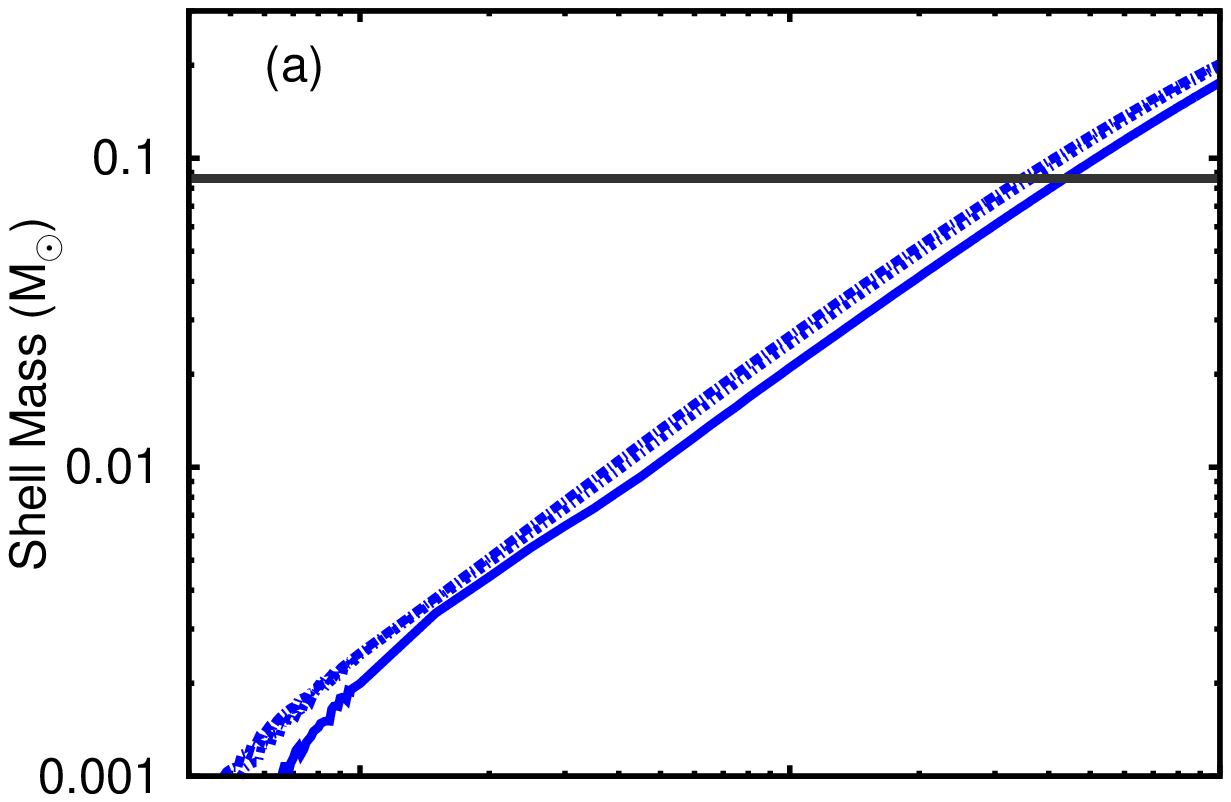}
\includegraphics[width=0.48\hsize]{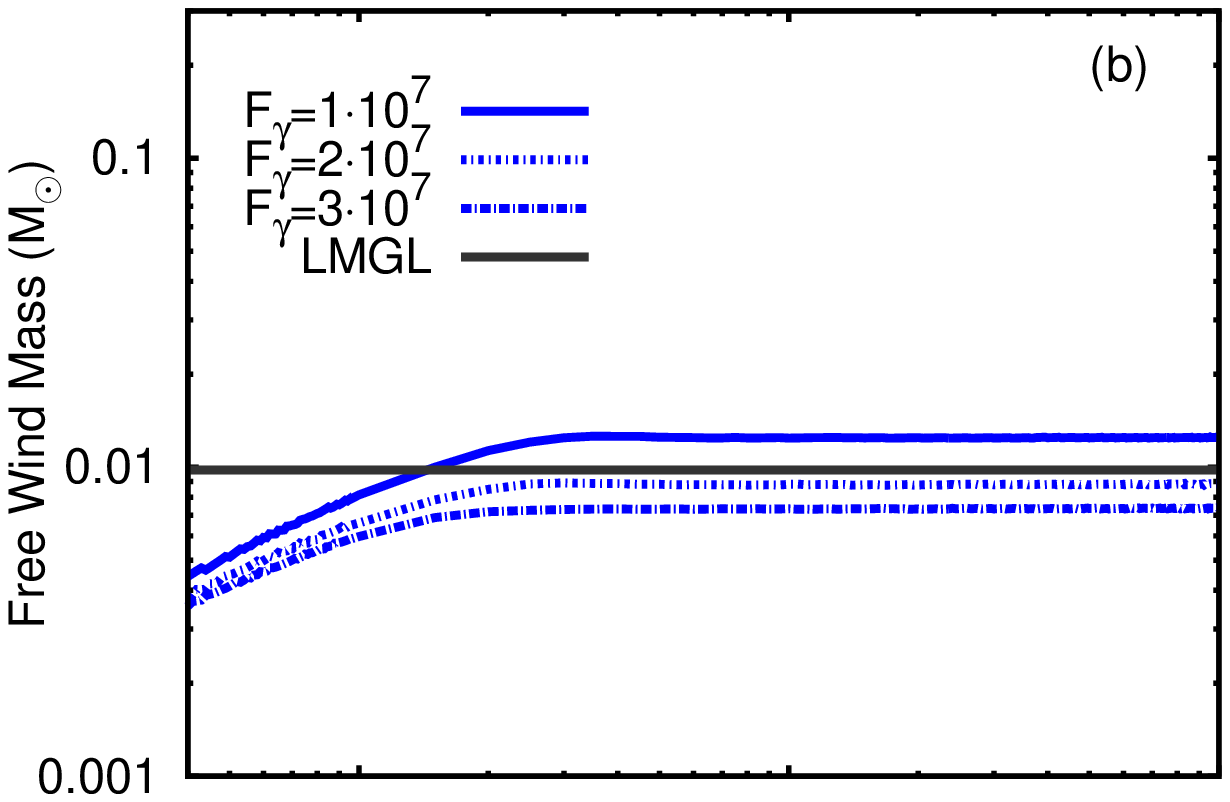}
\includegraphics[width=0.48\hsize]{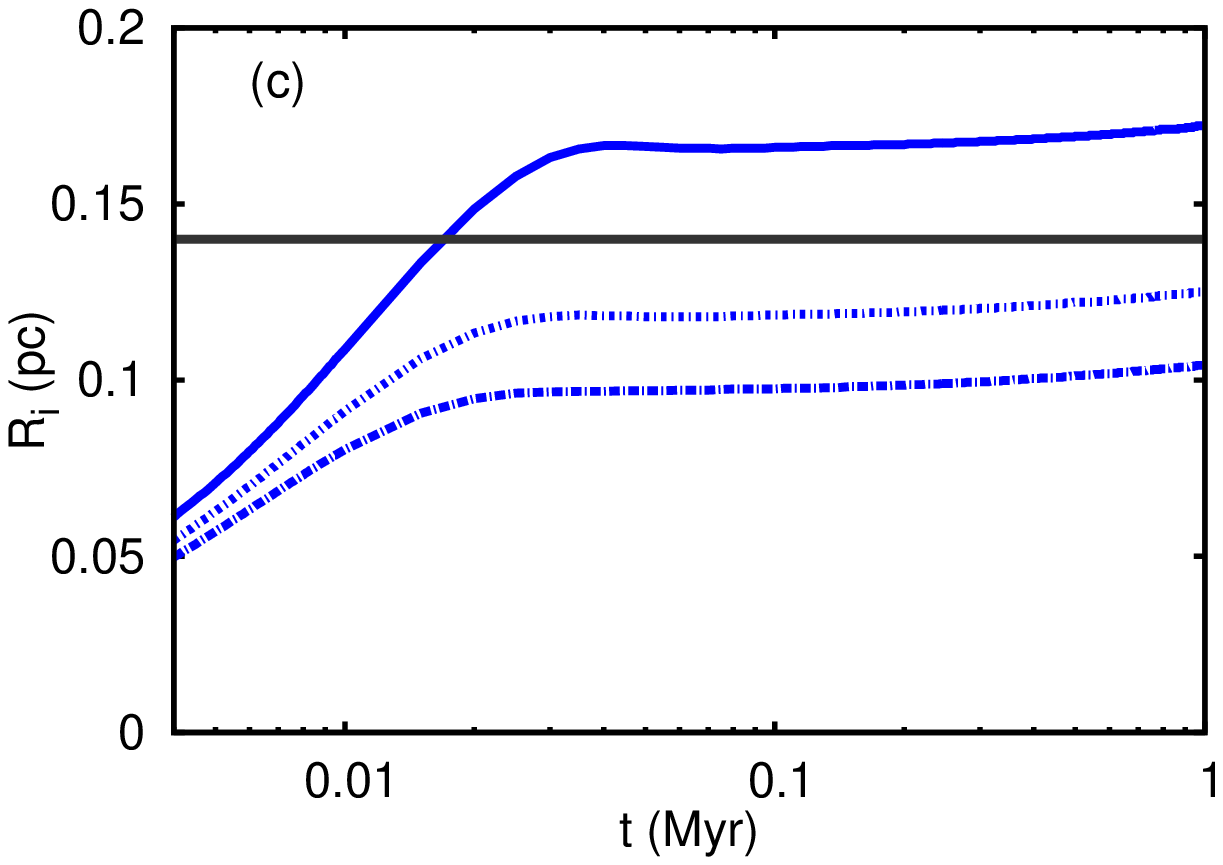}
\includegraphics[width=0.48\hsize]{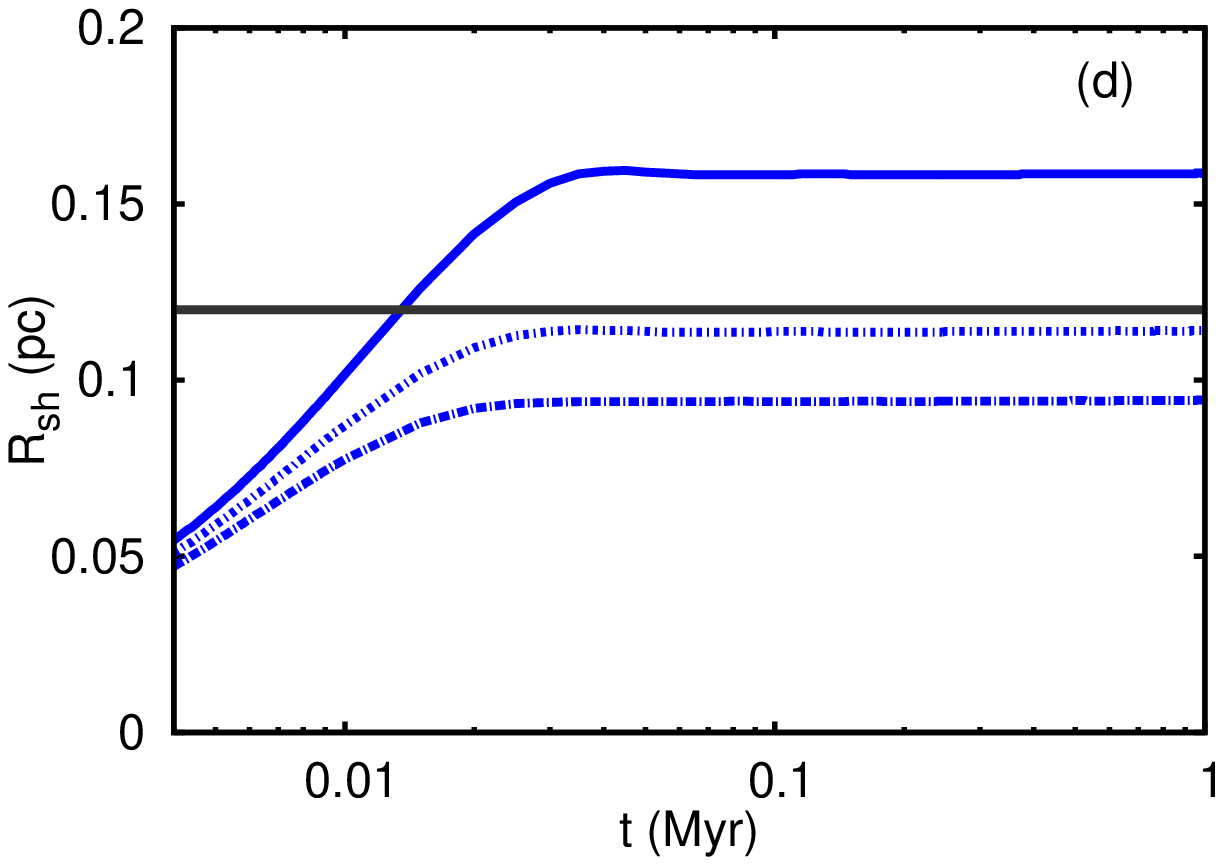}
\end{center}
\caption{
  \textbf{
  Time evolution of  the photoionization-confined shell around Betelgeuse for spherically symmetric simulations with three different ionizing fluxes.}
  \textbf{(a)} $M_{\mathrm{shell}}$; \textbf{(b)} mass in the freely-expanding wind, $M_{\mathrm{wind}}$; \textbf{(c)} $R_\mathrm{IF}$; \textbf{(d)} $R_{\mathrm{shell}}$.
  All calculated using wind parameters $\dot{M}=1.2\times10^{-6}\ M_{\odot}\,\mathrm{yr}^{-1}$ and $v_\mathrm{n}=14\ \mathrm{km}\,\mathrm{s}^{-1}$.
  The three blue curves are from simulations with different external ionizing fluxes, $F_\gamma$ ($\mathrm{cm}^{-2}\,\mathrm{s}^{-1}$).
  The black lines are plotted using data in table 2 of ref.~\cite{LeBMatGerEA12} ($M_{\mathrm{DT,CS}}$, $M_{r<r_1}$, $r_\mathrm{f}$ and $r_1$ and \textbf{a-d}, respectively), calibrated to match the 21\,cm H~\textsc{i} observations.
\label{fig:AOri_shell}
}
\end{figure}

The steady-state shell mass for Betelgeuse is $M_\mathrm{shell} = 1.0\ M_{\odot}$, for the parameter values\cite{LeBMatGerEA12}  $v_\mathrm{n}=14\ \mathrm{km}\,\mathrm{s}^{-1}$, $R_\mathrm{IF}\approx0.15$ pc, and $\dot{M}=1.2\times10^{-6}\ M_{\odot}\,\mathrm{yr}^{-1}$ (stellar mass-loss rate).
If $\dot{M}$ is larger (for example $3\times10^{-6}\ M_{\odot}\,\mathrm{yr}^{-1}$; ref.~\cite{MohMacLan12}) then $M_\mathrm{shell}$ increases accordingly.
In Fig.~\ref{fig:AOri_shell} we compare our model predictions for Betelgeuse to
a recent analysis\cite{LeBMatGerEA12} in which 21\,cm H~\textsc{i} observations were interpreted in the context of a detached shell model.
A photoionization-confined shell matches the observations well
for an external ionizing flux of $F_\gamma\approx2\times10^7\ \mathrm{cm}^{-2}\,\mathrm{s}^{-1}$.
Such a flux is found near the edge of old H~\textsc{ii} regions or within interstellar bubbles where diffuse ionizing photons constitute a large fraction of the total flux\cite{Rit05} 
 (Methods; Extended Data Fig.~4 and Supplementary Video 1 show results from this simulation).
The observed shell mass ($0.09\ M_{\odot}$) constrains its age to be $0.3 - 0.5$ Myr.
Betelgeuse's post-main-sequence lifetime is about 1 Myr, and so it will probably explode before the photoionization-confined shell attains its steady-state mass.

\begin{figure}
\begin{center}
\includegraphics[width=0.7\hsize]{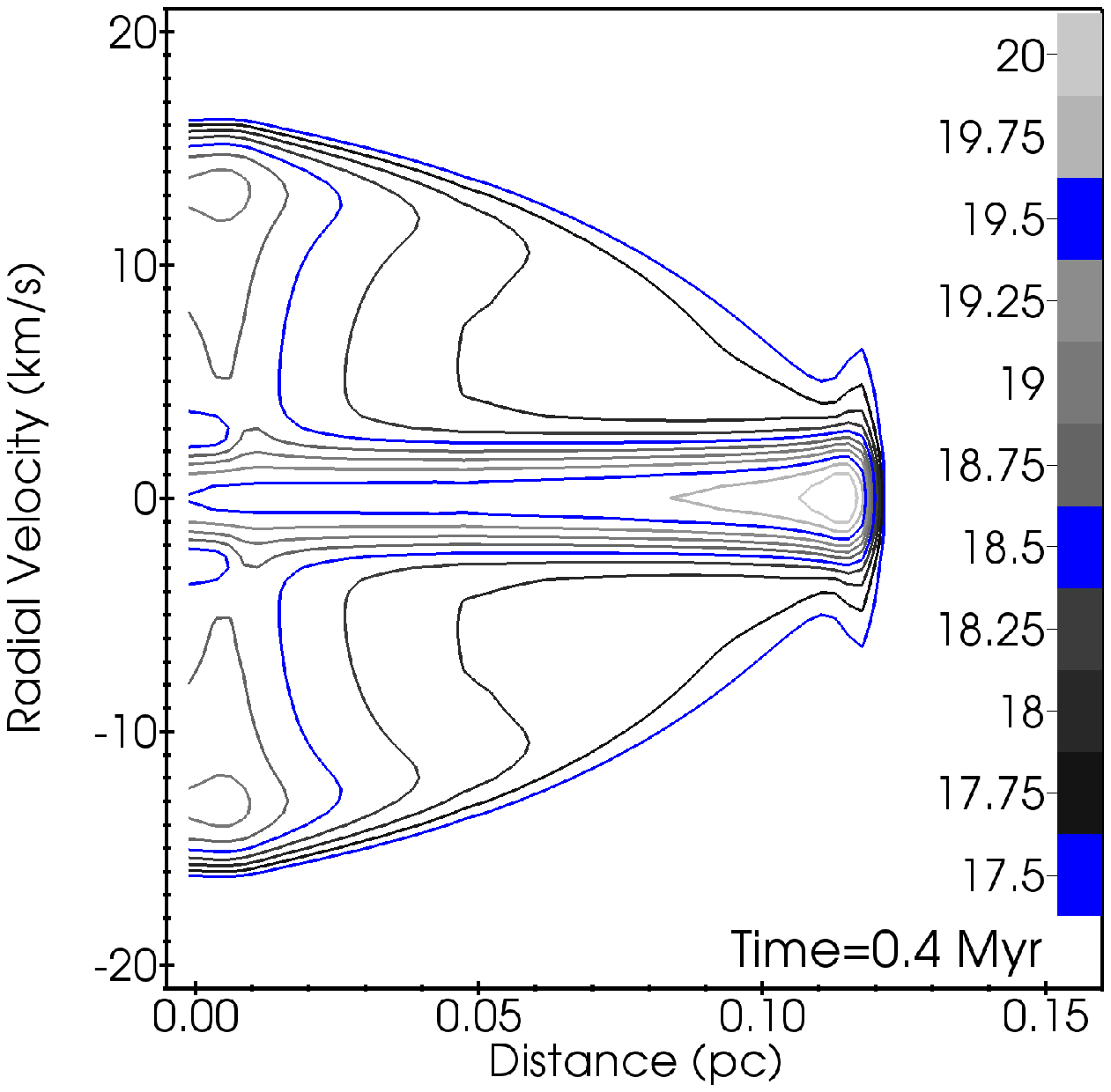}
\includegraphics[width=0.7\hsize]{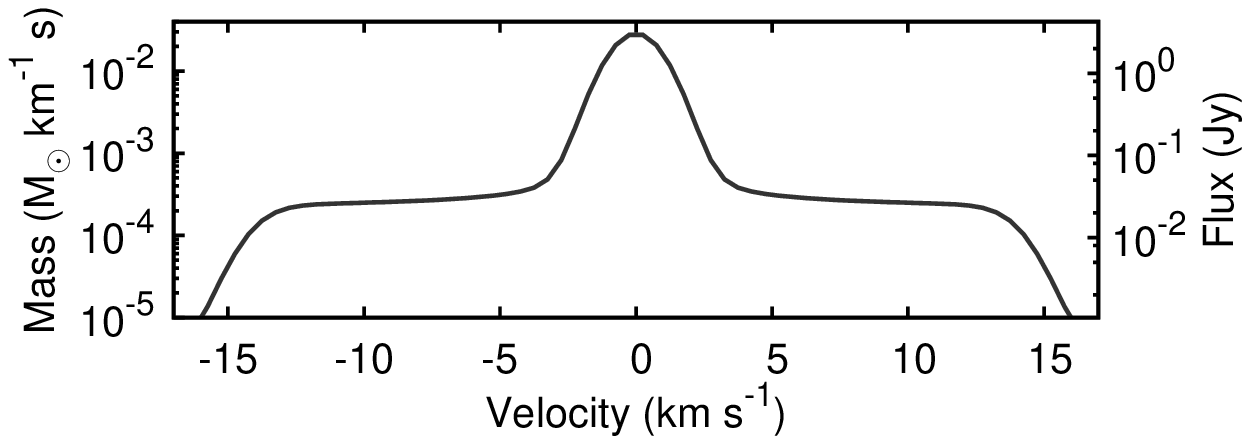}
\end{center}
\caption{
    \textbf{
    Simulated observations of neutral hydrogen in the photoionization-confined shell around Betelgeuse.}
   Observations are the output of a simulation using the models described in Fig.~\ref{fig:AOri_shell} with $F_\gamma=2\times10^{7}\ \mathrm{cm}^{-2}\,\mathrm{s}^{-1}$ at $t=0.4$ Myr, when $M_\mathrm{shell}=0.093\ M_{\odot}$.
   {\bf [Above]} Position-velocity diagram showing logarithmic contours of H~\textsc{i} column density as a function of projected distance from the star and radial velocity, in units of $\log_{10}$ (H~\textsc{i} atoms per cm$^2$ per $\mathrm{km}\,\mathrm{s}^{-1}$).
  The freely-expanding wind is seen red- and blueshifted by 14 $\mathrm{km}\,\mathrm{s}^{-1}$ (thermally broadened), and the almost static shell is at zero velocity and is limb-brightened at large radius.
  {\bf [Below]} Total spectrum of the H~\textsc{i} emission, assuming a distance of 200 pc and that the source is  unresolved and spherically symmetric (mass of H~\textsc{i} per unit velocity also shown).
  \label{fig:AOri_PV}
  }
\end{figure}

Further quantitative comparison at simulation time $t=0.4$ Myr is shown in Fig.~\ref{fig:AOri_PV}.
The H~\textsc{i} column density is plotted in a position-velocity diagram as a function of separation from Betelgeuse and line-of-sight velocity of the gas.
The blue- and red-shifted components of the freely expanding wind are at radial velocity $v_r\approx\pm14\ \mathrm{km}\,\mathrm{s}^{-1}$, and the shocked shell is centred on $v_r=0\ \mathrm{km}\,\mathrm{s}^{-1}$.
Our results again agree quantitatively with the data presented in ref.~\cite{LeBMatGerEA12} (fig.~11 therein), with the caveat that our simulations did not self-consistently determine the shell temperature (nor, consequently, the thermal broadening).

Although the neutral shell has been clearly seen observationally, a crucial prediction of our model is that the photoionization-confined shell should be surrounded by an accelerating ionized wind, emitting bremsstrahlung at radio wavelengths (with an emission measure of $\sim10-20$\ cm$^{-6}$\,pc; emission measure is the integral of the square of the electron number density along a line of sight) and Doppler-shifted nebular spectral lines.
Such a nebula has been detected for the red supergiant W\,26\cite{DouClaNegEA10,WriWesDreEA14}, but it will be orders of magnitude fainter around Betelgeuse because the latter's wind has much lower density.
The only imaging detections of Betelgeuse's gaseous circumstellar medium so far are the 21 cm H~\textsc{i} data and unexplained far-UV emission from the bow shock\cite{LeBMatGerEA12}.

The agreement of Betelgeuse's neutral shell with our photoionization-confined shell calculations is encouraging, but further work is required to investigate multidimensional effects such as the non-radial flows that are introduced by clumpy and asymmetric winds\cite{SmiHinRyd09}, the dynamical stability of the shocked shell, and the anisotropic external radiation fields\cite{MorJur83} (Methods).
Photoionization-confined shells may also be present around lower-mass red giants and stars on the asymptotic giant branch that have winds of comparable velocity, if they are located in a photoionized medium.

The main effect of a photoionization-confined shell is to confine much more gas (up to 80 times more (Methods)) close to a red supergiant than would  be expected from a freely expanding wind.
Simulations show that $\approx20-35$\% of the red supergiant wind is trapped in the photoionization-confined shell (Extended Data Fig.~3).
The shell mass is ultimately limited by the total mass shed by the star during the red supergiant phase of evolution, which is typically less than $20\ M_{\odot}$ at solar metallicity, and so we expect the most massive shells to have $M_\mathrm{shell}\approx(4-7)\ M_{\odot}$.

This has important implications for supernova/circumstellar-medium interactions because
ejected material from about 10\% of all core-collapse supernovae is observed to collide with dense circumstellar matter in the immediate vicinity of the exploding star\cite{Sma09}.
It is usually assumed that this dense matter is produced by short periods of extraordinarily high mass-loss rate ($\lesssim0.1\,M_{\odot}\,\mathrm{yr}^{-1}$) just before the star explodes\cite{SmiLiFolEA07, SmiFolBloEA08, SmiHinRyd09, YooCan10, FoxFilSkrEA13, MorMaeTadEA13}.
This gas is difficult to decelerate and confine close to the star hydrodynamically\cite{vMarLanAchEA06},
requiring a prompt explosion after an eruptive mass-loss event.
There is, however, no proven evolutionary link between eruptions and explosions (although ideas are being investigated\cite{QuaShi12,Mor14}).
The photoionization-confined shell scenario overcomes this timing problem because the wind \textit{is} decelerated effectively, 
allowing a fundamentally different interpretation of the lightcurves of some interacting supernovae.
For example, Betelgeuse's mass-loss rate was previously deemed too small to produce an interacting supernova\cite{SmiHinRyd09}, but this conclusion may need revision following the detection of its photoionization-confined shell\cite{LeBMatGerEA12}.

\begin{figure}
\begin{center}
\includegraphics[width=\hsize]{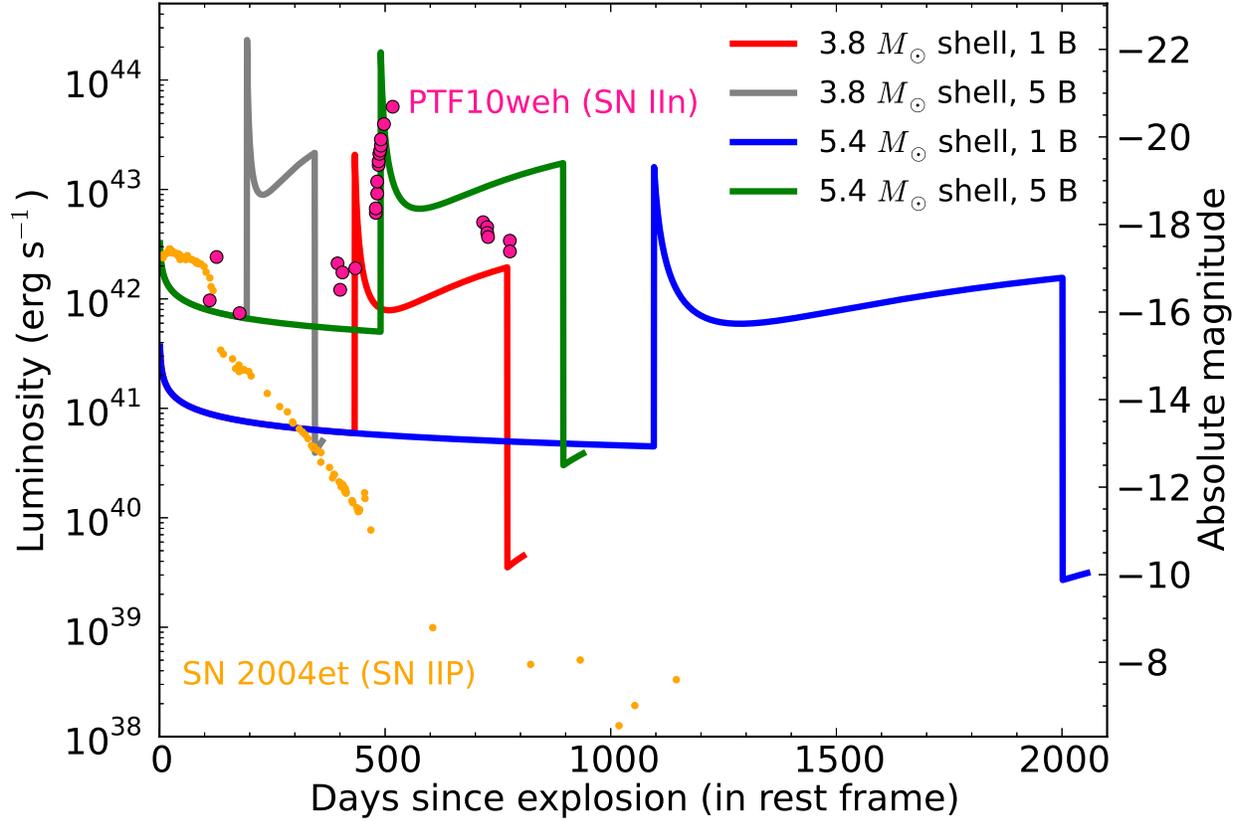}
\end{center}
\caption{
  \textbf{
  Predicted luminosity evolution of supernovae interacting with massive photoionization-confined shells, compared with observations of two core-collapse supernovae.}
  The shells are from simulations with $\dot{M}=10^{-4}\ M_{\odot}\,\mathrm{yr}^{-1}$, $v_\mathrm{n}=15\ \mathrm{km}\,\mathrm{s}^{-1}$, and either $F_\gamma=10^{14}\ \mathrm{cm}^{-2}\,\mathrm{s}^{-1}$ ($3.8\ M_{\odot}$ shell) or $F_\gamma=10^{13}\ \mathrm{cm}^{-2}\,\mathrm{s}^{-1}$ ($5.4\ M_{\odot}$ shell at larger radius).
  The shell grows for 0.2 Myr until the star has lost $20\ M_{\odot}$ of mass, and then it explodes.
  The supernova kinetic energy is either $10^{51}$ erg (equal to 1 Bethe (B)) or $5\times10^{51}$  erg (5 B), as indicated.
  Also plotted are R-band lightcurves of the interacting supernova PTF10weh (ref.~\cite{OfeSulShaEA14}) and the more typical supernova 2004et (ref.~\cite{KotMeiFarEA09}).
  \label{fig:lightcurve}
  }
\end{figure}

Results from a calculation of a radiative supernova blastwave expanding through the circumstellar medium of two of our most extreme models are plotted in Fig.~\ref{fig:lightcurve}.
The bolometric lightcurve rebrightens when the blastwave reaches the photoionization-confined shell, and remains at nearly constant luminosity until the shell is overrun by the shock.
The initial circumstellar medium interaction is strong enough to leave detectable signatures in supernova observations, and the later shell collision is even more luminous.
Two core-collapse supernova lightcurves are shown for qualitative comparison: SN 2004et (ref.~\cite{KotMeiFarEA09}) belongs to the most common (plateau) class, whereas SN PTF10weh (ref.~\cite{OfeSulShaEA14}) is an interacting supernova.
The lightcurve of PTF10weh was interpreted as a bright pre-supernova eruption ($0-500$ d.\ in Fig.~\ref{fig:lightcurve}) followed by a luminous supernova at $\sim500$ d.\ (ref.~\cite{OfeSulShaEA14}).
Fig.~\ref{fig:lightcurve} shows an alternative interpretation that fits the data well:  an explosion at 0 days followed by a collision of the supernova shock with a photoionization-confined shell at $\sim500$ d.
In this interpretation the supernova would produce broad spectral lines immediately after explosion.
Because pre-supernova eruptions eject lower velocity gas with narrower lines, this could be used to distinguish the two scenarios.

\section*{Bibliography}
\vspace{1cm}

\bibliographystyle{naturemag}


\begin{addendum}
 \item 
  JM and SM are grateful to P.~Kervella, T.~Le Bertre and G.~Perrin, the organisers of the Betelgeuse Workshop in Paris (Nov.~2012), where the ideas for this work were first developed.
  JM acknowledges funding from a fellowship from the Alexander von Humboldt Foundation and from the Deutsche Forschungsgemeinschaft priority program 1573, ``Physics of the Interstellar Medium''.
  SM acknowledges the receipt of research funding from the National Research Foundation (NRF) of South Africa.
  TJM is supported by Japan Society for the Promotion of Science Postdoctoral Fellowships for Research Abroad (26\textperiodcentered 51).
  HRN acknowledges funding from a NSF grant (AST-0807664).
  RK acknowledges support from STFC (ST/L000709/1).
The authors acknowledge the John von Neumann Institute for Computing for a grant of computing time on the JUROPA supercomputer at J\"ulich Supercomputing Centre.
 \item[Competing Interests] The authors declare that they have no
competing financial interests.
 \item[Correspondence] Correspondence and requests for materials
should be addressed to JM~(email: jmackey@astro.uni-bonn.de).
 \item[Author contributions]
  JM and SM had the original idea that Betelgeuse's static shell could be confined by external radiation.
  JM derived the analytic equations for the shell, and ran and analysed the spherically symmetric computations.
  VVG, DMAM, NL, and JM discussed the results in the context of recently discovered photoionized winds, which motivated many of the specific choices of parameters used.
  JM, SM, VVG, DMAM, HRN, NL interpreted Betelgeuse's shell in the context of our results.
  NL proposed that the shells could be relevant for interacting supernovae, and developed this idea with JM, RK and TJM.
  Figures were prepared by JM, SM, TJM, RK.
  All authors contributed to the writing of the manuscript.
\end{addendum}

\newpage

\newpage

\newpage

\newpage

\begin{methods}

\subsection{Analytic model for photoionization-confined shells}
\label{sec:theory}
Consider a steady-state, spherically symmetric wind around a star that loses mass at a rate $\dot{M}$.
The wind density and velocity are $\rho(r)$ and $v(r)$, respectively, as a function of distance from the star, $r$.
The number density of H atoms is $n_{\mathrm{H}} = X_\mathrm{H}\rho/m_\mathrm{p}$, where $X_\mathrm{H}$ is the mass fraction of H (the solar value\cite{AspGreSauEA09} is $X_\mathrm{H}=0.715$) and $m_\mathrm{p}$ is the proton mass.
For simplicity we assume that He remains neutral in the red supergiant wind and calculate only the ionization and recombination of H (introducing an error of at most 10\%).
The electron and proton number densities are then equal, $n_\mathrm{e}=n_\mathrm{p}=n_{\mathrm{H}}(1-y)$, where $y$ is the neutral fraction of H.
The recombination rate of H is $A = \alpha_\mathrm{B} n_e n_\mathrm{p} $ in units of $\mathrm{cm}^{-3}\,\mathrm{s}^{-1}$, where $\alpha_\mathrm{B}$ is the (temperature-dependent) case B recombination coefficient\cite{Hum94}, with numerical value $\alpha_\mathrm{B}\approx2.7\times10^{-13} \ \mathrm{cm}^3\,\mathrm{s}^{-1}$ for a photoionized gas temperature $T_i=10^4$ K.

The wind is externally photoionized by an isotropic radiation field with inward ionizing flux $F_\gamma$.
An ionization front forms at radius $R_\mathrm{IF}$, separating the neutral inner wind from the photoionized outer wind.
We use an approximate two-phase equation of state in which both ionized and neutral phases are isothermal with temperatures $T_\mathrm{i}$ and $T_\mathrm{n}\ll T_\mathrm{i}$, respectively (and associated sound speeds $a_\mathrm{i}$ and $a_\mathrm{n}\ll a_\mathrm{i}$).
Using $T_\mathrm{i}=10^4$ K implies $a_\mathrm{i}=11.1\ \mathrm{km}\,\mathrm{s}^{-1}$, which is typical for photoionized gas with Galactic chemical composition.
The isothermal approximation implies that the cooling time (and length) behind a shock is zero.
The internal shell structure from our simulations therefore does not capture the post-shock cooling region, but this does not strongly affect our conclusions because they are based primarily on pressure balance at the shell boundaries.
A non-zero cooling length means that photoionization-confined shells may be thicker in reality than in our models, but most of the mass will still be in the thin, dense, cooled layer, and so the shell's observable properties are probably very similar to what we predict using the isothermal approximation.

The wind terminal velocity through $R_\mathrm{IF}$ is $v_\mathrm{n}$.
Assuming that $v(r)=v_\mathrm{n}$ at all radii, we can integrate the number of recombinations in the wind from $r=\infty$ inwards to obtain $R_\mathrm{IF}$ as a function of the wind properties ($\dot{M}$ and $v_\mathrm{n}$) and ionizing flux $F_\gamma$.
This gives\cite{MorJur83}
\begin{align}
R_\mathrm{IF} &= \left(\frac{\alpha_\mathrm{B}}{3F_\gamma}\right)^{1/3}
      \left(\frac{X_\mathrm{H}\dot{M}}{4\pi v_\mathrm{n} m_\mathrm{p}} \right)^{2/3} \nonumber\\
       &= 0.018\ \mathrm{pc} 
  \left( \frac{\dot{M}}{10^{-4}\ M_{\odot}\,\mathrm{yr}^{-1}} \right)^{2/3}
  \left( \frac{F_\gamma}{10^{13}\ \mathrm{cm}^{-2}\,\mathrm{s}^{-1}} \right)^{-1/3} 
  \left( \frac{v_\mathrm{n}}{15\ \mathrm{km}\,\mathrm{s}^{-1}} \right)^{-2/3} \,.
\label{eqn:IFront_radius}
\end{align}

This equation assumes that photoionization has no hydrodynamic effect on the wind.
If $v_\mathrm{n}>2a_\mathrm{i}$ then the ionization front is R-type (rarefied), characterised by weak density and velocity changes across the front, with no associated shocks\cite{Kah54,Axf61}.
If $v_\mathrm{n}\leq 2a_\mathrm{i}$ then a D-type (dense) ionization front occurs, consisting of a shock propagating into the neutral wind, a shocked neutral shell and an ionization front\cite{Kah54,Axf61,MihMih84} (with a strong density jump between neutral and ionized gas) at larger radius.
The shock velocity relaxes to $v_\mathrm{n}$ once a steady state is reached.
If the shocked gas can cool efficiently then the shell can be very dense.

The ionized wind emerging from the D-type ionization front will accelerate to $30-40\ \mathrm{km}\,\mathrm{s}^{-1}$, according to the solution for a thermally driven isothermal wind\cite{Par58, MihMih84}.
In the absence of gravity (it is irrelevant at the radii we are considering) the steady-state equation for the radial velocity profile is slightly modified from the cited references, to
\begin{equation}
\left(\frac{v(r)}{a_\mathrm{i}}\right)^2 -2\ln \left(\frac{v(r)}{a_\mathrm{i}}\right) = 1+4\ln \left(\frac{r}{r_0}\right) \;.  \nonumber
\label{eqn:wind_acc}
\end{equation}
Here $r_0$ is the radius at which the flow passes through the sonic point.
Near $r=r_0$ the velocity scales as $\sqrt{r/r_0}$, and at large radius $v(r)\approx 2 a_\mathrm{i}\sqrt{\ln(r/r_0)}$.
Ionization fronts with diverging ionized gas flows, as is the case here, are D-critical\cite{Kah54,HenArtGar05}, for which ionized gas is accelerated to $a_\mathrm{i}$ within the ionization front itself.
This means that $r_0$ is coincident with the ionization front radius $R_\mathrm{IF}$.
We therefore set the velocity of the gas leaving the ionization front to $v_\mathrm{i}=a_\mathrm{i}$.

The photoionized wind evidently does not have constant velocity, so equation~(\ref{eqn:IFront_radius}) provides only an approximate estimate of $R_\mathrm{IF}$.
It should remain reasonably accurate because the velocity of the ionized flow remains within a factor of 2-3 of $v_\mathrm{n}$ for red supergiants, and the velocity dependence of $R_\mathrm{IF}$ is not strong.
Results of numerical tests in Extended Data Fig.~1 verify this.
$R_\mathrm{IF}$ scales as expected with $\dot{M}$ and $F_\gamma$, but is independent of $v_\mathrm{n}$ for $v_\mathrm{n}\leq2a_\mathrm{i}$.
Replacing $v_\mathrm{n}$ with $16\ \mathrm{km}\,\mathrm{s}^{-1}$ provides a good fit to the numerical results in this case, and  we make this substitution for the numerical results throughout the paper.

The shell is bounded upstream by a standing shock in the neutral wind with isothermal shock jump conditions 
$\rho_\mathrm{shell}a_\mathrm{n}^2 = \rho_\mathrm{n} v_\mathrm{n}^2$,
where $\rho_\mathrm{n}$ is the wind density just upstream of the shock radius $R_{\mathrm{shell}}$, and $\rho_\mathrm{shell}$ is the shell density.
The flow through the shell is subsonic and isothermal, and so the shell maintains a constant density.
Its outer boundary is the ionization front, at radius $R_\mathrm{IF}$.
Conservation of mass and momentum in the steady-state flow, together with the shock jump condition, gives the ratio
\begin{equation}
\frac{R_\mathrm{IF}^2}{R_{\mathrm{shell}}^2} = \frac{v_\mathrm{i}^2+a_\mathrm{i}^2}{v_\mathrm{i}v_\mathrm{n}} - \frac{a_\mathrm{n}^2}{v_\mathrm{n}^2} \;.
\label{eqn:shell_radius}
\end{equation}
For strong shocks (cold winds) the second term on the right hand side of equation~(\ref{eqn:shell_radius}) is small compared with the first, and so we discard it for the rest of the analysis.  The shell mass, $M_\mathrm{shell}$, is then 
\begin{align}
M_\mathrm{shell} &= \frac{\dot{M}}{3} R_\mathrm{IF} 
    \left(\frac{v_\mathrm{i}^2+a_\mathrm{i}^2}{v_\mathrm{i}a_\mathrm{n}^2}\right)
    \left(1 - \left[\frac{v_\mathrm{i}^2+a_\mathrm{i}^2}{v_\mathrm{i}v_\mathrm{n}}\right]^{-3/2}\right)
    \nonumber\\
    &=
    \frac{2\dot{M}}{3}\frac{a_\mathrm{i}}{a_\mathrm{n}^2} R_\mathrm{IF} 
    \left(1 - \left[\frac{v_\mathrm{n}}{2a_\mathrm{i}}\right]^{3/2}\right)
    \;, \label{eqn:shell_mass}
\end{align}
where the second line is obtained by setting $v_\mathrm{i}=a_\mathrm{i}$.
Note that $M_\mathrm{shell}\rightarrow0$ as $v_\mathrm{n}\rightarrow2a_\mathrm{i}$, because for larger velocities a shell-forming D-type ionization front is not possible.
$M_\mathrm{shell}$ is also sensitive to $a_\mathrm{n}$, and colder shells can become much more massive.
This is because $R_\mathrm{IF}$ and $R_{\mathrm{shell}}$ are independent of the shell density, whereas the shell density scales with $a_\mathrm{n}^{-2}$.

If we then assume that $R_\mathrm{IF}$ deviates little from the value predicted by equation~(\ref{eqn:IFront_radius}), we find that
\begin{align}
M_\mathrm{shell} & =  \left(\frac{\alpha_\mathrm{B}X_\mathrm{H}^2}{162\pi^2 m_\mathrm{p}^2}\right)^{1/3} 
    \frac{a_\mathrm{i}}{a_\mathrm{n}^2} \left(1 - \left[\frac{v_\mathrm{n}}{2a_\mathrm{i}}\right]^{3/2}\right)
    \dot{M}^{5/3}F_\gamma^{-1/3}  v_\mathrm{n}^{-2/3}
    \nonumber\\
    & = 9.2\ M_{\odot} 
    \left( \frac{\dot{M}}{10^{-4}\ M_{\odot}\,\mathrm{yr}^{-1}} \right)^{5/3}
    \left( \frac{F_\gamma}{10^{13}\ \mathrm{cm}^{-2}\,\mathrm{s}^{-1}} \right)^{-1/3}
    \;, \label{eqn:shell_mass_num} \nonumber
\end{align}
where we have used the following numerical values: $v_\mathrm{n}=15\ \mathrm{km}\,\mathrm{s}^{-1}$, $a_\mathrm{i}=11.1\ \mathrm{km}\,\mathrm{s}^{-1}$, $a_\mathrm{n}=0.81\ \mathrm{km}\,\mathrm{s}^{-1}$, $\alpha_\mathrm{B}=2.7\times10^{-13} \ \mathrm{cm}^3\,\mathrm{s}^{-1}$, and $X_\mathrm{H}=0.7154$.
If we ignore the scaling of $R_\mathrm{IF}$ with $v_\mathrm{n}$ (as argued above), then the factor of $v_\mathrm{n}^{-2/3}$ should be replaced with $(16\ \mathrm{km}\,\mathrm{s}^{-1})^{-2/3}$.
Results from numerical simulations in Extended Data Fig.~2 show that this equation provides a very good fit to the steady-state shell mass.
To summarise the approximations, we have assumed that
\begin{enumerate}
\item the flow has relaxed to steady state with spherical symmetry,
\item the wind speed is low enough to permit a D-type ionization front ($v_\mathrm{n}\leq 2a_\mathrm{i}$),
\item the neutral wind is ram-pressure dominated,
\item the neutral and ionized gas phases are both isothermal,
\item the ionization front is treated as a discontinuity with an outflow velocity $v_\mathrm{i}\approx a_\mathrm{i}$, and
\item the ionization front is at the same radius it would be if the shell did not exist.
\end{enumerate}

The shell mass can be much larger than the freely-expanding wind mass, $M_\mathrm{wind}=\dot{M}R_\mathrm{IF}/v_\mathrm{n}$, that would otherwise occupy the circumstellar medium.
We obtain
\begin{equation}
\frac{M_\mathrm{shell}}{M_\mathrm{wind}} = \frac{2}{3} 
    \left(1 - \left[\frac{v_\mathrm{n}}{2a_\mathrm{i}}\right]^{3/2}\right)
    \frac{v_\mathrm{n}a_\mathrm{i}}{a_\mathrm{n}^2}
    \;.  \nonumber \label{eqn:mass_ratio}
\end{equation}
This ratio is independent of $\dot{M}$ and $R_\mathrm{IF}$ and, for $v_\mathrm{n}=14\ \mathrm{km}\,\mathrm{s}^{-1}$ and the sound speeds given above, it is $\sim 80$.

The shell mass is ultimately limited by the mass shed during the red supergiant phase of evolution, which is typically less than $20\ M_{\odot}$ at solar metallicity.
In many cases, the final steady-state shell mass is not reached, as seen from Extended Data Fig.~2 where extreme shells have steady-state masses $M_\mathrm{shell} >100\ M_{\odot}$.
The timescale for shell growth is
\begin{equation}
\tau_\mathrm{shell} \equiv \frac{M_\mathrm{shell}}{f\dot{M}} = 
  \frac{1}{f}\frac{2a_\mathrm{i}}{3 a_\mathrm{n}^2} R_\mathrm{IF} 
  \left(1 - \left[\frac{v_\mathrm{n}}{2a_\mathrm{i}}\right]^{3/2}\right) \;,
    \label{eqn:shell_time}
\end{equation}
where $f$ is the fraction of the wind mass retained in the shell.
The time evolution of $M_\mathrm{shell}$ for Betelgeuse (with $f\approx0.2$) and a more extreme model (with $f\approx0.35$) are plotted in Extended Data Fig.~3.
Their shell growth times are $\tau_\mathrm{shell}=4.2$ and 0.21 Myr, respectively.
The simulations show that $f$ is approximately constant until the shell reaches $1/3$ to $1/2$ of its steady-state mass.

\subsection{Radiation-hydrodynamics simulations}
We use the radiation hydrodynamics code \textsc{pion}\cite{MacLim10,Mac12} for our numerical simulations, with spherical symmetry and, consequently, one (radial) degree of freedom, and using a uniform, fixed grid in the radial coordinate.
\textsc{Pion} uses a finite-volume discretisation of the equations of hydrodynamics, solved with an explicit time-integration scheme that is accurate to second order in time and space.
The non-equilibrium ionization and recombination of hydrogen are coupled to the hydrodynamics using algorithm 3 in ref.~\cite{Mac12}.
The simple two-temperature isothermal equation of state means that gas temperature depends only on the neutral fraction of hydrogen, $y$, according to
$T(y) = T_\mathrm{n} + (T_\mathrm{i}-T_\mathrm{n})(1-y)$,
so that $T(y=1)=T_\mathrm{n}$ (the cold neutral gas temperature) and $T(y=0)=T_\mathrm{i}$ (the hot ionized gas temperature).

The ionizing photon spectrum is taken to be that of a late O-type star, which emits relatively few photons capable of ionizing helium\cite{MarSchHil05}.
We consider a black-body spectrum (with temperature $T_r=3\times10^4$ K) between the ionization potentials of H$^0$ and He$^0$ (13.6 and 24.4 eV, respectively), we assume He remains neutral at all times, and that the radiation is isotropic and coming from infinity.
The only effect of the chosen spectrum when using the isothermal equation of state is to set the thickness of the ionization front, which has no material effect on the shell properties.

The simulation domain is set so that $R_\mathrm{IF}/13 \leq r \leq 5R_\mathrm{IF}$ (with $R_\mathrm{IF}$ from equation~\ref{eqn:IFront_radius}).
This ensures that the inner (inflow) and outer (outflow) boundaries do not affect the solution in any way.
Five thousand, one hundred and twenty grid zones were used for the final results, and this was tested to ensure numerical convergence.
The simulations were run for at least 5 growth timescales (using equation~\ref{eqn:shell_time} with $f=0.25$), and were checked to ensure that a steady state had been reached.

A plot of gas density, temperature, velocity, and wind fraction (a tracer with value 1 in the stellar wind and 0 in the interstellar medium (ISM)) is shown in Extended Data Fig.~4 after a red supergiant wind has been expanding for 10,000 years (Supplementary Video 1 shows the time evolution).
From small to large radius, it shows the freely-expanding wind, the thin photoionization-confined shell at $r\approx0.08$ pc bounded by the ionization front, the accelerated photoionized wind region in $0.09\ \mathrm{pc}\lesssim r\lesssim0.17$ pc, the wind termination shock at $r\approx0.17$\,pc, the contact discontinuity at $r\approx0.2$ pc, and a forward shock in the ISM at $r\approx0.23$ pc.
The shocked shell at the wind-ISM contact discontinuity cannot trap the ionization front and so remains fully photoionized, even for the low ionizing flux of $F_\gamma=2\times10^{7}\ \mathrm{cm}^{-2}\,\mathrm{s}^{-1}$ used for this simulation.
The flux must be decreased by a further factor of ten before the wind-ISM interface can trap the ionization front and prevent the formation of a photoionization-confined shell.

\subsection{The circumstellar medium around Betelgeuse}
As one of the two closest red supergiants to Earth, we have a uniquely detailed view of Betelgeuse's complex circumstellar medium:
its arc-shaped bow shock at a radius $r\approx0.35$ pc from the star\cite{NorBurCaoEA97, MohMacLan12, DecCoxRoyEA12},
the mysterious bar-shaped structure lying in the star's path just beyond the bow shock\cite{NorBurCaoEA97, UetIzuYamEA08, MacMohNeiEA12, DecCoxRoyEA12},
and the newly discovered, almost static neutral shell\cite{LeBMatGerEA12} closer to the star at $r\approx0.12-0.15$\,pc.
This shell cannot be explained as a stellar eruption or wind variation because it is static
(that is,~some external force has decelerated the wind),
nor as hydrodynamic confinement by the ISM because this occurs at larger radius at the bow shock.

Previous hydrodynamic simulations\cite{MohMacLan12,DecCoxRoyEA12} have shown that the bow shock around Betelgeuse should be quite massive, and also unstable.
This is in apparent contradiction to the \textit{Herschel} observations that show rather smooth arcs.
When a red supergiant wind is photoionized, the ionized part of the wind accelerates to $>30\ \mathrm{km}\,\mathrm{s}^{-1}$, and the bow shock becomes more stable than when the wind is neutral\cite{MeyGvaLanEA14}.
A photoionized bow shock may therefore fit the observations better than a neutral one.
The ISM magnetic field may also be able to suppress instabilities sufficiently to agree with observations\cite{VanDecMel14}.
The multiple arcs may be a projection effect from undulations in the shock surface; whatever the correct explanation, there is no indication that this  arc-shaped structure is anything other than a bow shock.

It was suggested that the bow shock may be associated with H~\textsc{i} emission\cite{LeBMatGerEA12}.
In our model, the bow shock should be photoionized, unless its densest region at the apex can self-shield sufficiently to allow it to partially recombine.
In this case the solid angle of the recombined region (seen from the star) cannot be too large, or else it would reduce the ionizing flux reaching the photoionization-confined shell.
The best evidence for  bow shock H~\textsc{i} emission is fig.\ 9 in ref.~\cite{LeBMatGerEA12}, which presents data summed over a large radial velocity range (unlike the data for the photoionization-confined shell) and could represent foreground or background ISM emission.
The \textit{GALFA} H~\textsc{i} data show no local maximum of emission at the bow shock\cite{DecCoxRoyEA12} (albeit with low spatial resolution).
Both studies\cite{LeBMatGerEA12,DecCoxRoyEA12} note that confusion with foreground and background gas along the line of sight is a significant issue in the data reduction, and so the evidence for this detection is weaker than for the emission associated with the photoionization-confined shell.

\subsection{Anisotropy of photoionization-confined shells}

NML Cyg, W26, and Betelgeuse are the three best-observed red supergiants with photoionized winds.
They display a range of morphologies, probably arising from the anisotropy of the ionizing radiation field.
For example, NML Cyg is illuminated only from one side and so the ionized part of its wind is bow shaped\cite{MorJur83}.
Betelgeuse's shell, by contrast, appears roughly spherical and so its irradiation must be more isotropic (the shell may be somewhat elongated in one direction\cite{LeBMatGerEA12}).
The H$\alpha$ ring around W26\cite{WriWesDreEA14} suggests that it too is irradiated from all sides, although the radio nebula has some asymmetry\cite{DouClaNegEA10}.
These examples point to a range of possible photoionization-confined shell shapes and masses.
The spherical case allows the most massive shells to form, because it has no non-radial flows.
For the asymmetric case, non-radial flows in the shocked shell cannot be much faster than the wind speed or the ionized gas sound speed, so it follows from advection timescales that even a completely one-sided shell should approximately double the mass of circumstellar gas near the star.
These are the extreme cases, so all photoionization-confined shells will increase the circumstellar mass by a factor of between $\sim2$ and $\sim80$ (see above).

\subsection{Source of the ionizing radiation} 
An O star at a distance of 100\,pc and with an ionizing photon luminosity of $L_\gamma=2.4\times10^{49}$ s$^{-1}$ will provide a flux of $F_\gamma=2\times10^7\ \mathrm{cm}^{-2}\,\mathrm{s}^{-1}$, if there is no absorption along the line of sight.
In projection Betelgeuse is certainly closer to $\lambda$ Ori than this.
This would, however, provide a directed radiation field and not an isotropic one.
An isotropic field can arise from the diffuse ionizing photons which pervade H~\textsc{ii} regions and superbubbles, and which are particularly important near H~\textsc{ii} region borders\cite{Rit05,WilHen09}.

The diffuse field is produced primarily by radiative recombinations directly to the ground state of H, emitting a photon with $h\nu>13.6$ eV.
The emission rate is\cite{Ost89} $\alpha_1=1.58\times10^{-13}$ cm$^{3}$\,s$^{-1}$ for $T=10^4$ K.
In equilibrium, the radiation intensity approaches the source function\cite{RybLig79} $S\equiv j/\alpha$, where $j= \alpha_1 n_e n_p /4\pi = \alpha_1 n_\mathrm{H}^2 (1-y)^2 /4\pi$ is the emissivity per unit volume and solid angle, and $\alpha= n_\mathrm{H} y \sigma_0$ is the absorption per unit length, where $\sigma_0=6.3\times10^{-18}$ cm$^{-2}$ is the threshold ionization cross-section of hydrogen.
For highly ionized gas, $1-y\approx1$, so $S\approx 2.0\times10^3 n_\mathrm{H}/y \ \mathrm{cm}^{-2}\,\mathrm{s}^{-1}\,\mathrm{sr}^{-1}$.
Using $n_\mathrm{H}=1\,\mathrm{cm}^{-3}$ and $y=5\times 10^{-4}$, the flux crossing a surface in one direction is $F_\gamma = \pi S = 1.3\times10^{7}\ \mathrm{cm}^{-2}\,\mathrm{s}^{-1}$.
This is comparable to the required flux, so our model is indeed viable if Betelgeuse is located near the edge of an H~\textsc{ii} region.
Possible evidence for this is the linear bar-like structure upstream from the bow shock\cite{NorBurCaoEA97}, which is interpreted as either a relic of a previous mass-loss phase of Betelgeuse\cite{MacMohNeiEA12} or as an interstellar density discontinuity\cite{DecCoxRoyEA12}, and so could be the shell at the edge of an H~\textsc{ii} region that Betelgeuse will soon encounter.
Alternatively, Betelgeuse may be within (or at the border of) the Orion-Eridanus Bubble\cite{BroHarBur95}, a hot bubble of ionized gas along the line-of-sight towards Orion.
In this case the same arguments apply except that both $n_\mathrm{H}$ and $y$ are lower than in a H~\textsc{ii} region.


\subsection{Lightcurve calculation}
Bolometric lightcurves are obtained by a method based on ref.~\cite{MorMaeTadEA13}.
We assume that the region shocked by the supernova forward and reverse
shocks forms a thin dense shell because of the efficient radiative
cooling\cite{CheFra94}.
This is confirmed to be a good approximation by numerical
radiation hydrodynamics simulations\cite{MorMaeTadEA13}.
The circumstellar medium wind density (proportional to $\dot{M}/v_\mathrm{n}$) for the models presented in Fig.~4 in the main text is similar to that in ref.~\cite{MorMaeTadEA13}, and so our assumption that the shock is radiative is valid.
Two-dimensional hydrodynamic simulations of supernovae interacting with circumstellar shells\cite{vMarSmiOwoEA10} also showed that radiative cooling is efficient and produced comparable lightcurves to the analytic method used here.

The evolution of the shocked, dense shell is simply governed by 
the conservation of momentum using these assumptions\cite{MorMaeTadEA13}, and so we solve this conservation equation numerically.
The density structure of the homologously expanding supernova ejecta is assumed to have two components.
The outer and inner supernova ejecta densities are assumed to be proportional to
$r^{-n}$ and $r^{-\delta}$, respectively.
Following the result of a numerical simulation of a red supergiant explosion\cite{MatMcK99}, we adopt $n=12$ and $\delta=0$,
although the lightcurves are not very sensitive to these choices.
We show models with the supernova ejecta mass of 15 $M_\odot$ and kinetic energies of $10^{51}$ erg and $5\times 10^{51}$ erg in Fig.~4.
The finite speed of light has not been taken into account, so the lightcurve features are sharper than for a real observation.
Also, the observations plotted on this figure have no bolometric corrections, and so the comparison is only indicative.

The Thomson scattering  optical depth of the photoionization-confined shells in the models are less than unity, so we neglect the effect of the shell opacity on the light curve.
We assume that 50\% of the available kinetic energy is converted to radiation because of the efficient radiative cooling in the shocked dense shell.
The fraction is uncertain and could be smaller because of, for example, multidimensional instabilities\cite{MorBliTomEA13}.
The reduction in the fraction results in the reduction of the bolometric luminosity but the lightcurve shapes remain the same.

Recent comparisons between observations and calculations of interacting supernova lightcurves\cite{MorMaeTadEA14} concluded that most progenitors had high mass-loss rates in the decades before explosion.
Our work is not in conflict with this conclusion because the real constraint is not on the mass-loss rate but on the circumstellar medium density as a function of radius from the progenitor star.
The further step of inferring a mass-loss rate for the progenitor assumes that the circumstellar medium is expanding at a constant rate and that none of it is decelerated by any environmental effects (radiative or hydrodynamic).
All previous calculations that sought to constrain the mass-loss history of the progenitor have made a similar assumption\cite{CheFra94, FoxFilSkrEA13}.
A consequence of the existence of photoionization-confined shells is that this assumption does not always hold, and that environmental effects can decelerate and confine the wind much closer to the star than previously thought\cite{vMarLanAchEA06}.

\end{methods}

\section*{Bibliography}
\vspace{1cm}

\newpage

\begin{figure}
\begin{center}
\includegraphics[width=0.48\hsize]{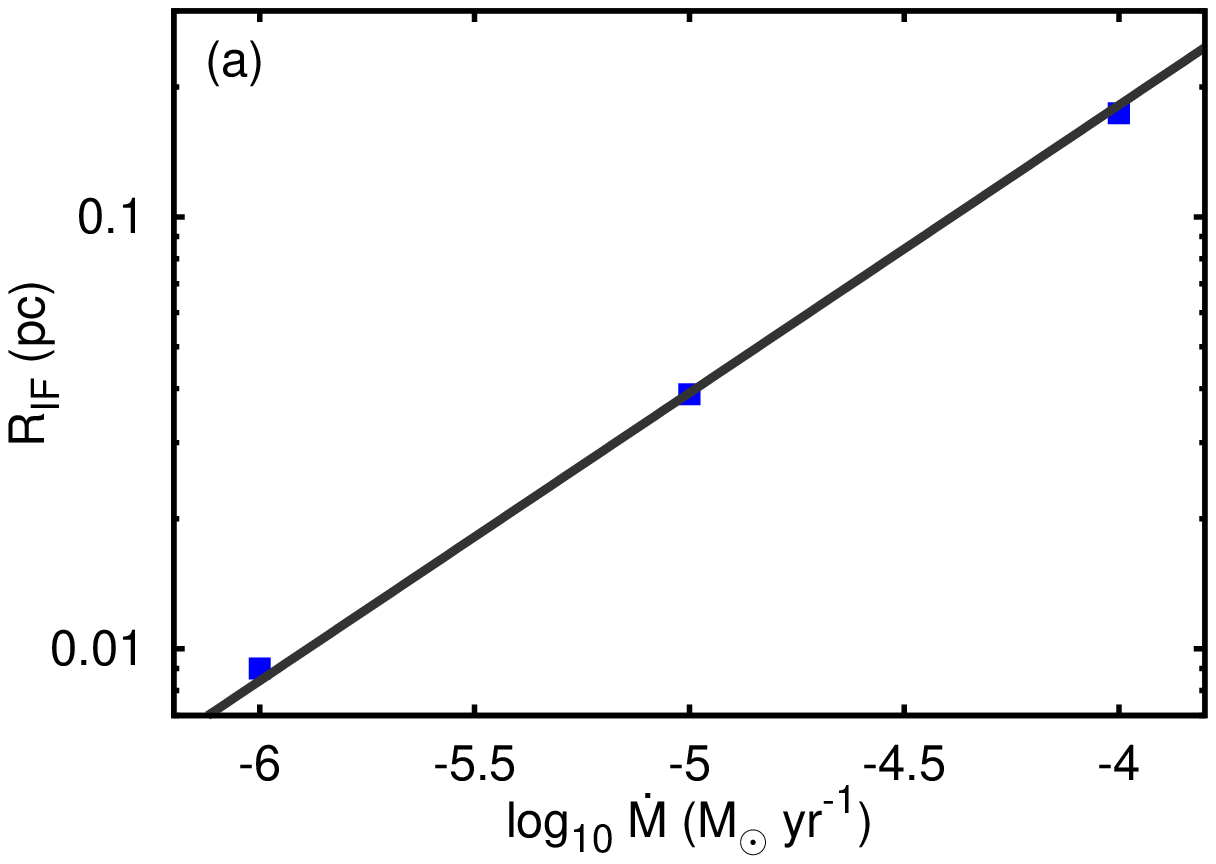}
\includegraphics[width=0.48\hsize]{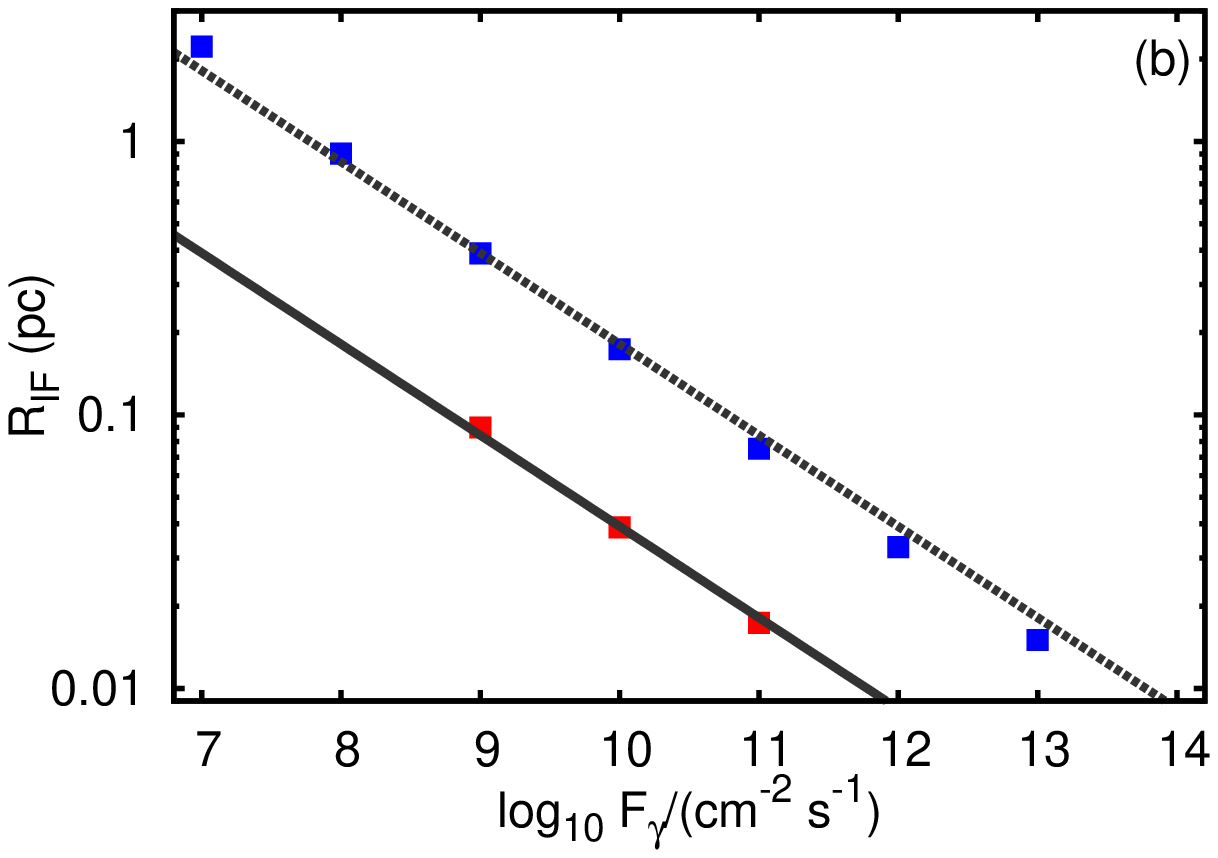}
\includegraphics[width=0.48\hsize]{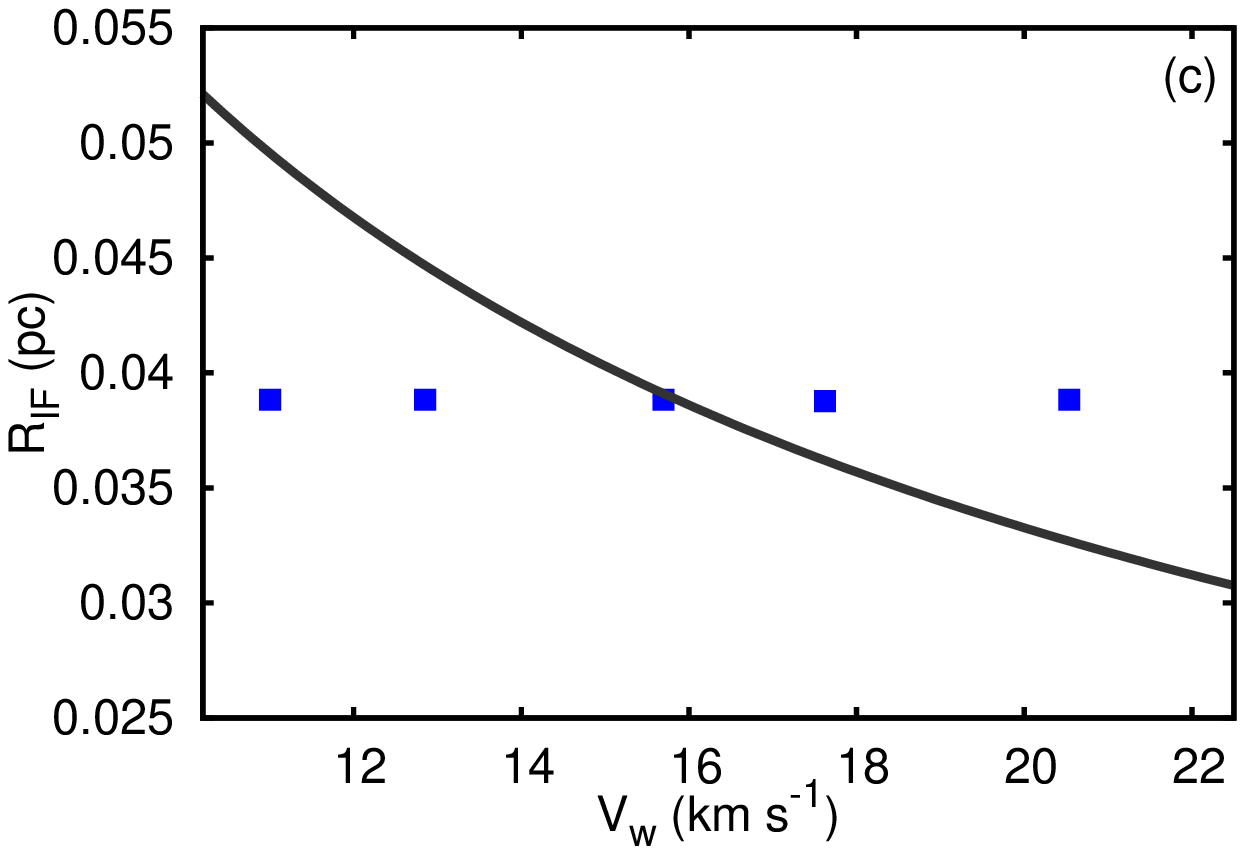}
\end{center}
\caption{
  {\bf Extended Data Figure 1\quad
  Dependence of the photoionization-confined shell radius on the properties of the stellar wind and external ionizing radiation.}
  The panels plot $R_\mathrm{IF}$ as a function of mass-loss rate, $\dot{M}$ \textbf{(a)}, external ionizing photon flux, $F_\gamma$ \textbf{(b)}, and wind velocity, $v_\mathrm{n}$ \textbf{(c)}.
  Data points are from spherically symmetric radiation hydrodynamics simulations and black lines are from equation~(\ref{eqn:IFront_radius}).
  In  \textbf{(a)} the fixed parameters are $v_\mathrm{n}=15\,\mathrm{km}\,\mathrm{s}^{-1}$ and $F_\gamma=10^{10}\ \mathrm{cm}^{-2}\,\mathrm{s}^{-1}$; in  \textbf{(b)} they are $v_\mathrm{n}=15\ \mathrm{km}\,\mathrm{s}^{-1}$ and either $\dot{M}=10^{-4}\ M_{\odot}\,\mathrm{yr}^{-1}$ (blue points) or $\dot{M}=10^{-5}\ M_{\odot}\,\mathrm{yr}^{-1}$ (red points); and in  \textbf{(c)} they are $\dot{M}=10^{-5}\ M_{\odot}\,\mathrm{yr}^{-1}$ and  $F_\gamma=10^{10}\ \mathrm{cm}^{-2}\,\mathrm{s}^{-1}$.
  }
  \label{fig:IFradius}
\end{figure}

\newpage

\begin{figure}
\begin{center}
\includegraphics[width=0.48\hsize]{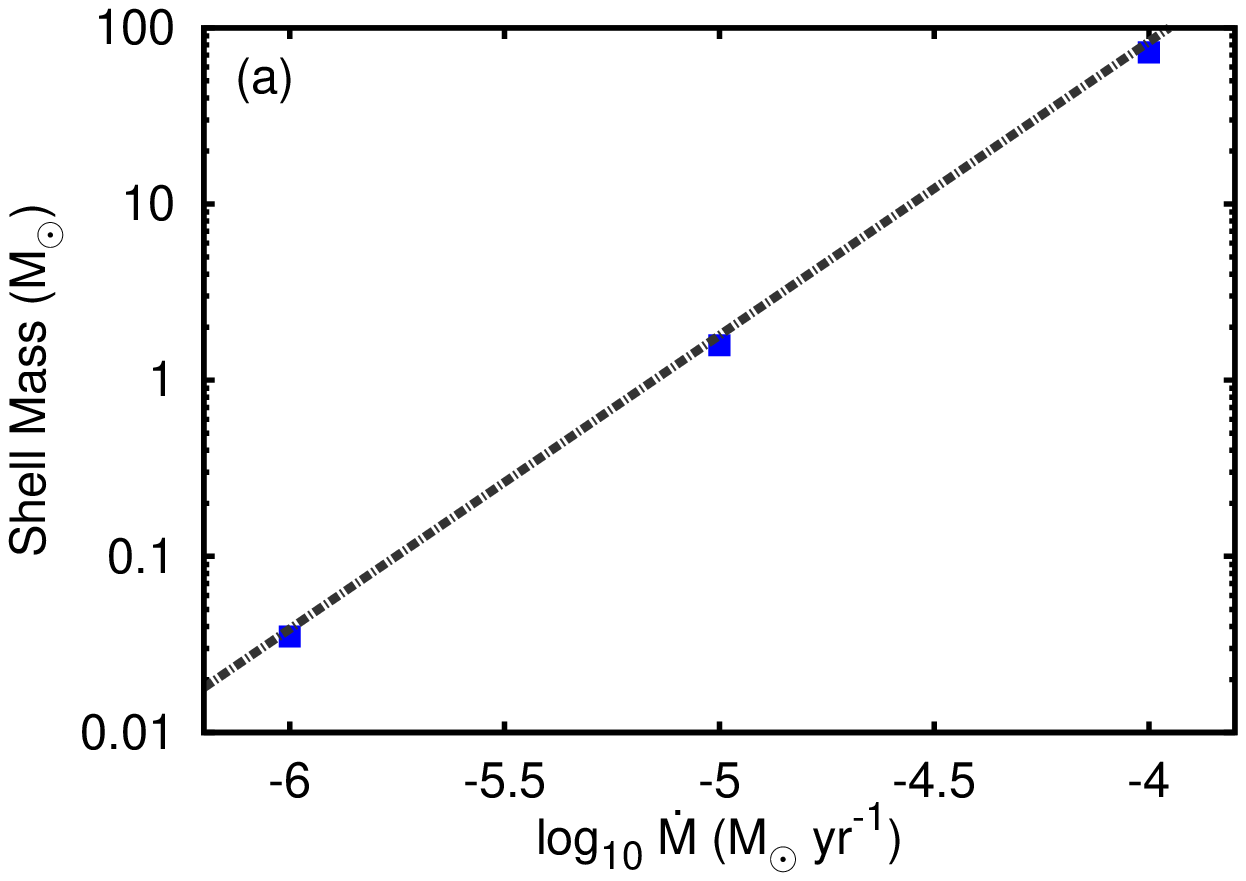}
\includegraphics[width=0.48\hsize]{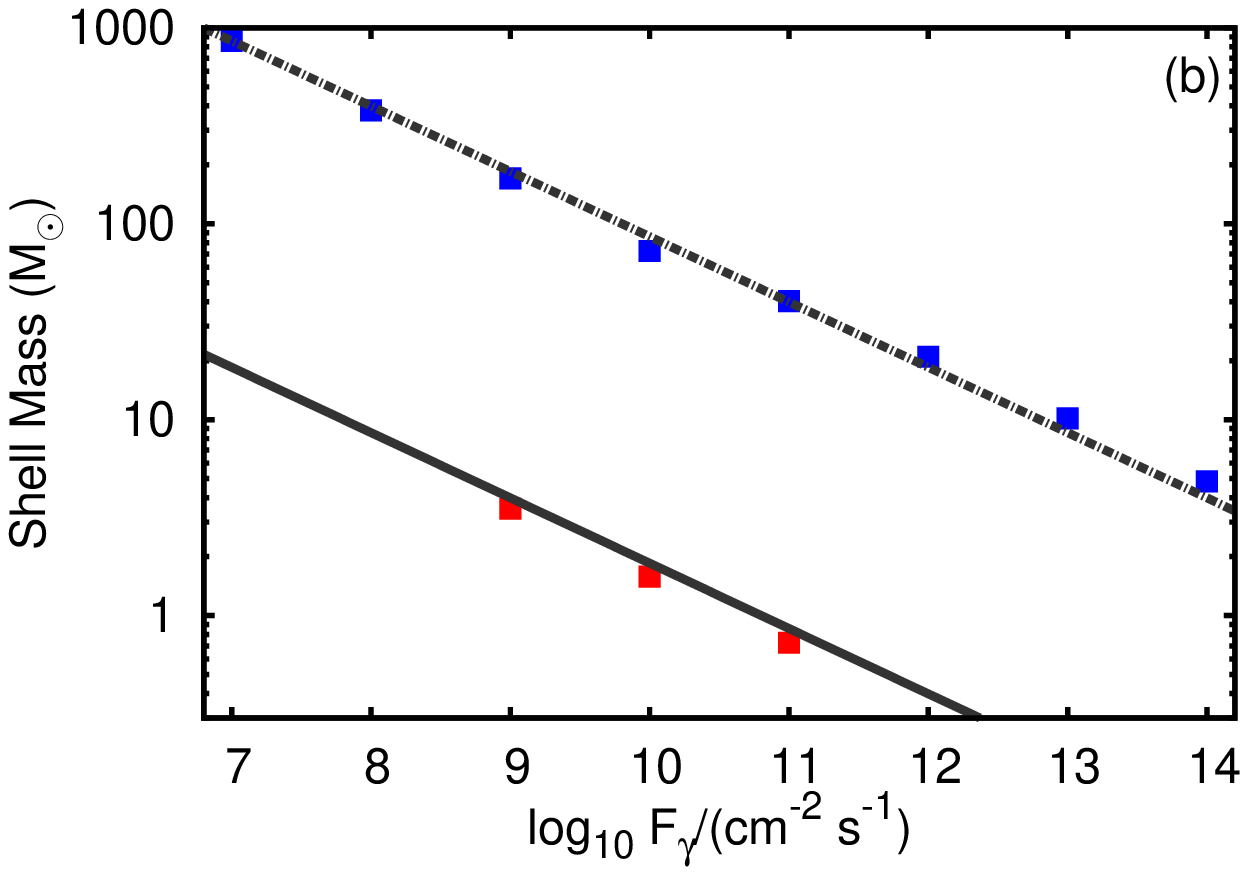}
\includegraphics[width=0.48\hsize]{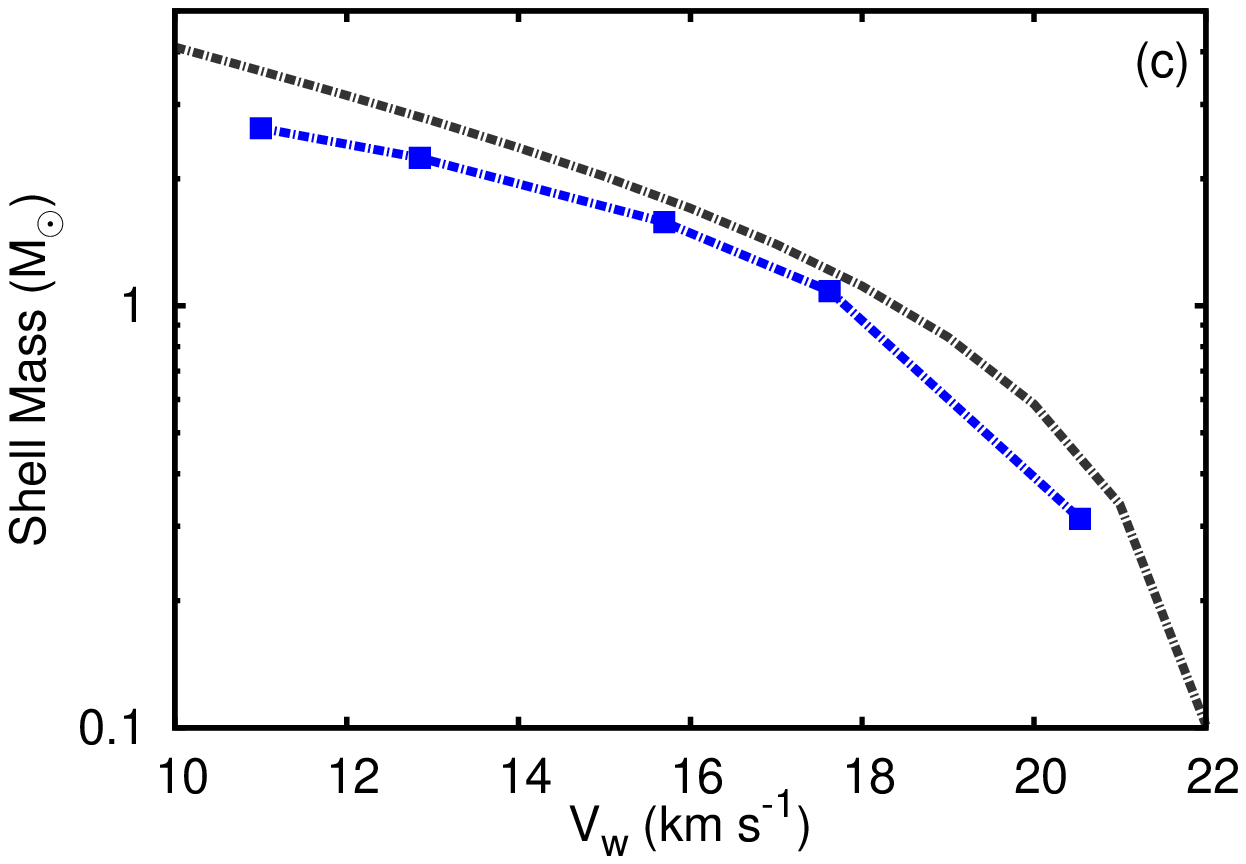}
\end{center}
\caption{
  {\bf Extended Data Figure 2\quad
  Dependence of the photoionization-confined shell mass on the properties of the stellar wind and external ionizing radiation.}
  The panels plot $M_\mathrm{shell}$ as a function of $\dot{M}$ \textbf{(a)}, $F_\gamma$ \textbf{(b)}, and $v_\mathrm{n}$ \textbf{(c)}.
  Data points are steady-state masses from spherically symmetric radiation hydrodynamics simulations and black lines are from equation~(\ref{eqn:shell_mass}).
  Again, in  \textbf{(a)} the fixed parameters are $v_\mathrm{n}=15\,\mathrm{km}\,\mathrm{s}^{-1}$ and $F_\gamma=10^{10}\,\mathrm{cm}^{-2}\,\mathrm{s}^{-1}$; in  \textbf{(b)} they are $v_\mathrm{n}=15\,\mathrm{km}\,\mathrm{s}^{-1}$ and either $\dot{M}=10^{-4}\,M_{\odot}\,\mathrm{yr}^{-1}$ (blue points) or $\dot{M}=10^{-5}\,M_{\odot}\,\mathrm{yr}^{-1}$ (red points); and in  \textbf{(c)} they are $\dot{M}=10^{-5}\,M_{\odot}\,\mathrm{yr}^{-1}$ and  $F_\gamma=10^{10}\,\mathrm{cm}^{-2}\,\mathrm{s}^{-1}$.
  }
\label{fig:shell_mass}
\end{figure}

\newpage

\begin{figure}
\begin{center}
\includegraphics[width=0.48\hsize]{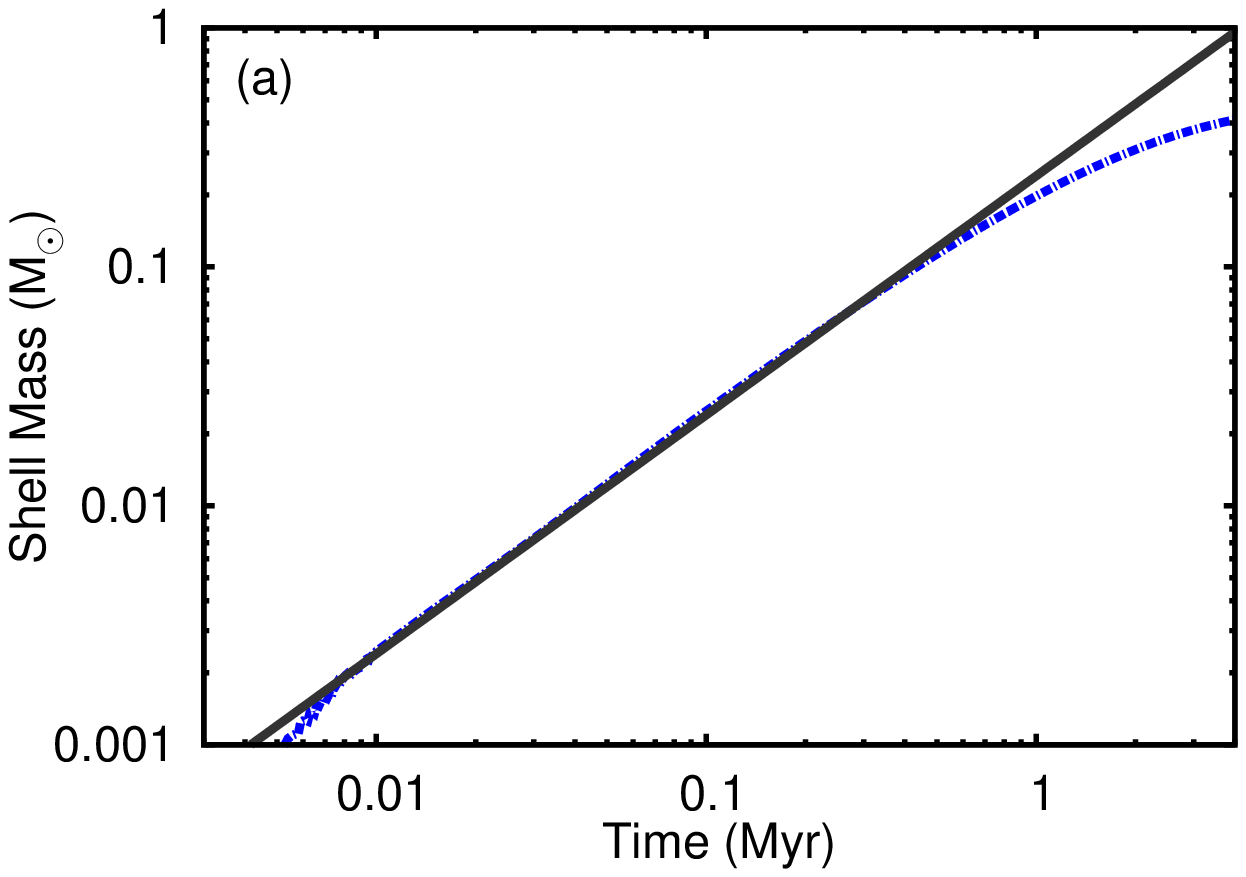}
\includegraphics[width=0.48\hsize]{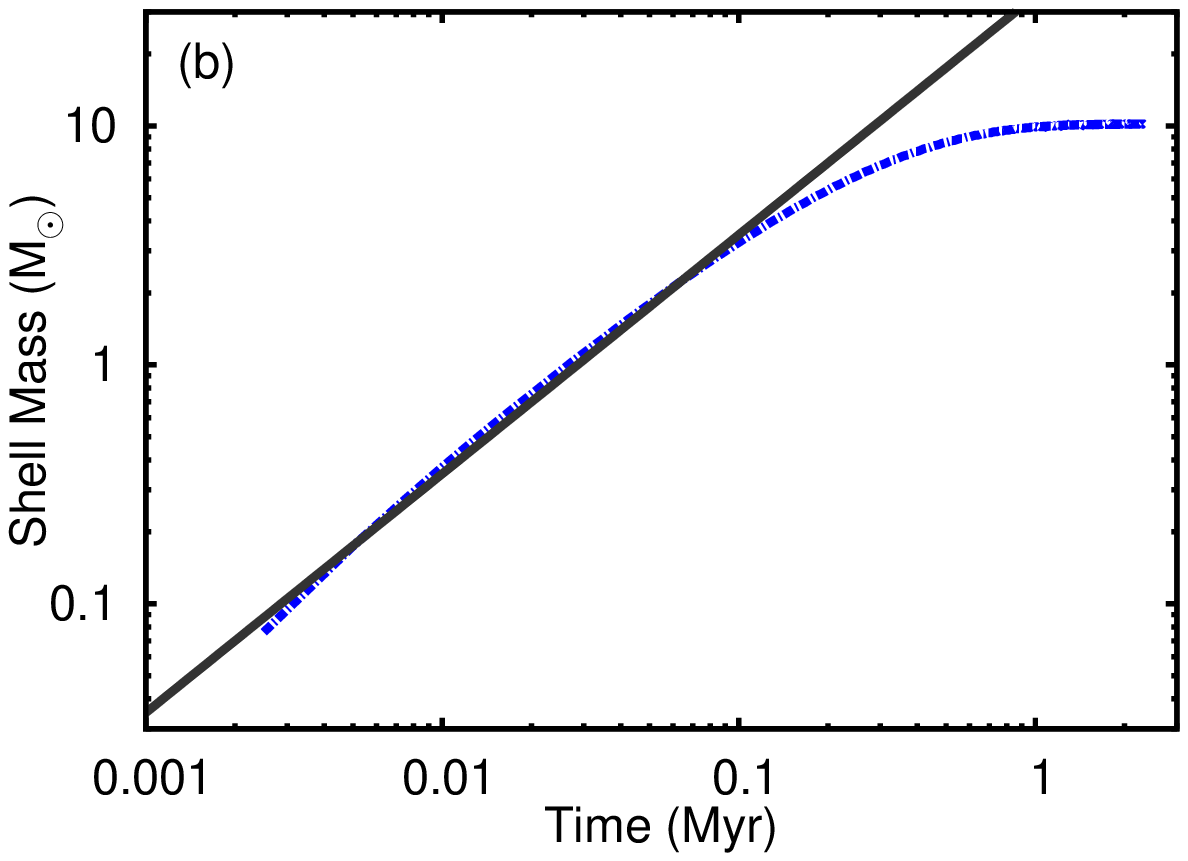}
\end{center}
\caption{
  {\bf Extended Data Figure 3\quad
  Growth of shell mass, $M_\mathrm{shell}$, as a function of time for two different photoionization-confined shell simulations.}
  The shell accumulates mass linearly with time until it begins to saturate at about one $1/3-1/2$ of its final mass.
  The solid line shows $M=0.2\dot{M}t$ in panel \textbf{(a)} and $M=0.35\dot{M}t$ in panel \textbf{(b)}.
   \textbf{(a)}, Photoionization-confined shell appropriate for Betelgeuse, with $\dot{M}=1.2\times10^{-6}\,M_{\odot}\,\mathrm{yr}^{-1}$, $v_\mathrm{n}=14\,\mathrm{km}\,\mathrm{s}^{-1}$, and $F_\gamma=2\times10^{7}\,\mathrm{cm}^{-2}\,\mathrm{s}^{-1}$.
   \textbf{(b)}, More extreme model with $\dot{M}=10^{-4}\,M_{\odot}\,\mathrm{yr}^{-1}$, $v_\mathrm{n}=15\,\mathrm{km}\,\mathrm{s}^{-1}$, and $F_\gamma=10^{13}\,\mathrm{cm}^{-2}\,\mathrm{s}^{-1}$.
  }
\label{fig:shell_growth}
\end{figure}

\newpage

\begin{figure}
\begin{center}
\includegraphics[width=\hsize]{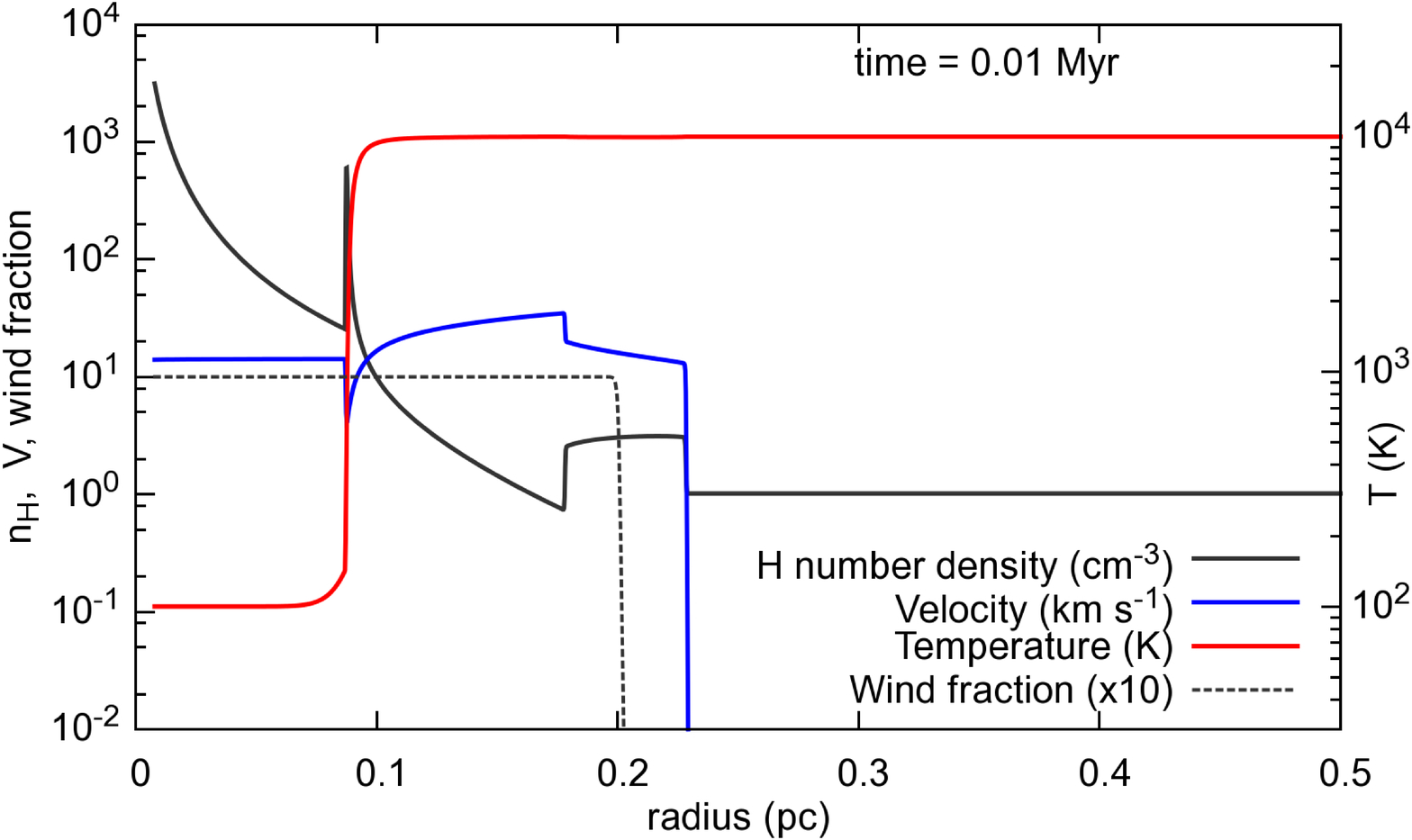}
\end{center}
\caption{
  {\bf Extended Data Figure 4\quad
  Structure of the circumstellar medium around Betelgeuse from a spherically symmetric radiation hydrodynamics simulation.}
  Hydrogen number density, gas velocity, temperature, and wind fraction are plotted as a function of distance from the star after 0.01 Myr of evolution.
  The wind fraction equals 1 in the wind and equals 0 in the ISM.
  The photoionization-confined shell is still very thin and has low mass at this early time, and the fully-ionized ISM interface at $r=0.2$ pc shows that the expanding wind drives a forward shock and a reverse shock.
  Supplementary Information contains a video showing an animation of the time evolution.
  }
\label{fig:radial_plot}
\end{figure}

\end{document}